\documentclass[preprint,authoryear,11pt]{elsarticle}

\usepackage{graphics}
\usepackage{graphicx}
\graphicspath{{Figures/}}
\usepackage{epsfig}
\usepackage{subfigure}
\usepackage{amssymb}
\usepackage{amsthm}
\usepackage{bm}
\usepackage{amsmath}
\usepackage{color}
\usepackage{geometry}
\usepackage{setspace}
\allowdisplaybreaks
\usepackage{enumerate}
\usepackage{booktabs}

\usepackage{hyperref}
\hypersetup{
	colorlinks=true,       
}

\biboptions{}

\journal{International Journal of Solids and Structures}

\begin{document}

\begin{frontmatter}



\title{Voltage-controlled topological interface states for bending waves
  \\in soft dielectric phononic crystal plates}


\author[1]{Yingjie Chen}
\author[2]{Bin Wu\corref{cor1}}
\ead{bin.wu@nuigalway.ie}
\author[1,2]{Michel Destrade}
\author[1,3]{Weiqiu Chen}

\cortext[cor1]{Corresponding author at: School of Mathematical and Statistical Sciences, University of Galway, University Road, Galway, Ireland.}
\address[1]{Key Laboratory of Soft Machines and Smart Devices of Zhejiang Province \\ and Department of Engineering Mechanics,\\ Zhejiang University, Hangzhou 310027, P.R. China;\\[6pt]}
\address[2]{School of Mathematical and Statistical Sciences, University of Galway, University Road, Galway, Ireland; \\[6pt]}
\address[3]{Soft Matter Research Center, Zhejiang University, Hangzhou 310027, China; }

\begin{abstract}
	
The operating frequency range of passive topological phononic crystals is generally fixed and narrow, limiting their practical applications. To overcome this difficulty, here we design and investigate a one-dimensional soft dielectric phononic crystal (PC) plate system with actively tunable topological interface states via the mechanical and electric loads. 
We use nonlinear electroelasticity theory and linearized incremental theory to derive the governing equations. 
First we determine the nonlinear static response of the soft dielectric PC plate subjected to a combination of axial force and electric voltage.
Then we study the motion of superimposed incremental bending waves. 
By adopting the Spectral Element Method, we obtain the dispersion relation for the infinite PC plate and the transmission coefficient for the finite PC plate waveguide. 
Numerical results show that the low-frequency topological interface state exists at the interface of the finite phononic plate waveguide with two topologically different elements. 
By simply adjusting the axial force or the electric voltage, an increase or decrease in the frequency of the topological interface state can be realized. 
Furthermore, applying the electric voltage separately on different elements of the PC plate waveguide is a flexible and smart method to tune the topological interface state in a wide range. 
These results  provide guidance for designing soft smart wave devices with low-frequency tunable topological interface states.

\end{abstract}

\begin{keyword}
Topological phononic crystal \sep active tunability\sep low-frequency interface state\sep dielectric elastomer \sep material nonlinearity



\end{keyword}

\end{frontmatter}




\underline{\underline{}}\section{Introduction}

Dielectric elastomers (DEs) are a type of smart materials that respond to rapidly to electric stimulus and develop large deformations. 
DEs have attracted enormous attention from academia and industry alike, due to excellent characteristics such as high energy density, high fracture toughness and light weight, and promising potential in artificial muscles, soft robotics, actuators, and energy harvesters  \citep{carpi2011dielectric, anderson2012multi, zhao2014harnessing}. 

Recently, it has also been shown that applying an electric field offers an effective approach to manipulating  acoustic/elastic waves in DEs via the induced finite deformations. 
Based on nonlinear electroelasticity theory and associated incremental theory \citep{dorfmann2006nonlinear, dorfmann2010electroelastic}, many investigations have been conducted to study superimposed infinitesimal waves in DEs subjected to the external mechanical and electric loads. 
For an infinite soft electroactive hollow tube with axial pre-stretch and axial electric field, \citet{su2016propagation}  analyzed the non-axisymmetric wave propagation characteristics. 
\citet{shmuel2016manipulating}  employed the stable Hybrid Matrix Method to investigate the manipulation of flexural waves in DE films controlled by an axial force and voltage. 
\citet{galich2016manipulating} studied both the propagation of pressure and shear waves in DEs under the action of electric stimuli. 
Based on the State Space Method, \citet{wu2017guided} presented a theoretical analysis of guided circumferential wave propagation in soft electroactive tubes under inhomogeneous electromechanical biasing fields, and found it could be used for ultrasonic non-destructive testing. 
\citet{wu2020nonlinear} studied nonlinear finite deformations and superimposed axisymmetric wave  in a functionally graded soft electroactive tube, when it is subjected to mechanical and electric biasing fields.  \citet{ziser2017experimental} showed experimentally that the flexural wave mode in a DE film can be tuned by voltage, and that the wave velocity can be slowed down. 
This setup was also modelled theoretically by \citet{broderick2020electro}.
 \citet{jandron2018numerical} demonstrated the tunable effect of electric load on the linearized wave propagation in infinite periodic composite DEs, based on Finite Element Method simulations.

Phononic crystals (PCs), which are essentially artificial periodic composites, have attracted intensive interests because of their outstanding characteristics in steering acoustic/elastic waves. 
Ascribed to the Bragg scattering \citep{kushwaha1993acoustic}, local resonance \citep{liu2000locally} or inertial amplification \citep{yilmaz2007phononic} mechanisms, the existence of a band gap (BG) is the most important feature possessed by PCs, wherein the propagation of acoustic/elastic wave is forbidden. Due to the exotic BG and the dispersive pass band properties, PCs can be applied to realize peculiar wave behaviors, such as negative refraction \citep{feng2006refraction,zhang2004negative}, cloaking \citep{zhang2011broadband,chen2017broadband} and one-way propagation \citep{fleury2014sound,chen2019tunable}. 

To achieve active control of wave propagation, many studies have been carried out to design PCs with wide tunable BGs, especially by using electric stimuli to tune the BGs in soft DE PCs. 
For the first time, \citet{shmuel2012band} investigated the electrostatically tunable BGs of incremental shear waves propagating perpendicular to the neo-Hookean ideal DE periodic laminates by the transfer matrix method. \citet{galich2017shear} re-checked the problem studied by \citet{shmuel2012band} and found that the BGs are not affected directly by the electric load for the shear waves propagating perpendicular to the layers in the neo-Hookean ideal DE laminates, which corrected the conclusion of \citet{shmuel2012band}. \citet{shmuel2013electrostatically} analyzed the propagation characteristics of incremental anti-plane shear waves in finitely extensible Gent DE fiber composites, and provided the first accurate demonstration of electrostatically tunable BGs in the DE composites. 
In addition, \citet{getz2017voltage} showed that the complete BGs of a soft dielectric fiber composite can be tuned by  electric voltage, due to the resulting changes in geometry and physical properties of the structure.
\citet{getz2017band} designed PC plates composed of two DE phases, and achieved voltage-controlled BGs, which can be used for active waveguides and isolators.
By adjusting the axial force and electric voltage applied to one-dimensional (1D) PC cylinders made of DE materials, \citet{wu2018tuning}  studied active tunability of superimposed longitudinal wave propagation. For a periodic compressible DE laminate, \citet{chen2020effects} shed light on the influence of pre-stress and electric stimuli on the nonlinear response and small-amplitude longitudinal and shear wave propagation behaviors. 
For more details on tunable and active PCs, the interested readers are referred to a recent review paper by \citet{wang2020tunable}.

Inspired by the concept of topological interface state in electronic systems, attention has been devoted in recent years to topological PC systems with particular topologically protected interface or edge states, which are unidirectional and immune to backscattering \citep{xiao2015geometric, ma2019topological}. 
The topological invariants, named Berry phase \citep{zak1989berry} for two-dimensional (2D) systems and Zak phase \citep{atala2013direct} for 1D systems, play an important role in characterizing the topological properties of band structures for PCs. 
Mimicking the quantum Hall effect, a first type of 2D topological PCs breaks the time-reversal symmetry with gyroscopes \citep{nash2015topological}, time-modulated materials \citep{chen2019mechanical}, or external flow fields \citep{khanikaev2015topologically}. 
The unidirectional topologically protected interface state in this type of topological PCs was  observed experimentally by  \citet{fleury2014sound}. 
A second type of 2D topological PCs breaks the spatial-inversion symmetry while conserving time-reversal symmetry, which support the pseudospin-dependent edge states and are named as quantum spin Hall topological insulators \citep{brendel2018snowflake, zhang2017experimental}. 
Because of the spin-orbit mechanism, quantum spin Hall topological PCs may feature forward and backward edge states by relying on appropriate polarization excitation \citep{yu2018elastic}. 
A third avenue is to break the inversion or mirror symmetry of 2D topological PCs, where the quantum valley Hall effect provides topologically protected interface states between two parts with opposite valley vortex states \citep{pal2017edge, wang2020tailoring}. For 1D topological systems, the topological transition process can be achieved by breaking spatial symmetry.
Hence, topological interface states were observed in waveguides composed of base elements with different topological properties \citep{xiao2015geometric, yin2018band}. 

However, for topological PCs made of passive materials, the working frequency range of topological interface/edge modes is narrow and fixed. 
In particular, the topological interface state in 1D systems usually emerges at a single frequency transmission peak in the overlapped BG. 
Therefore, actively tunable topological PCs are designed to possess a wider operating frequency range in practical applications. 
\citet{zhou2020actively} realized tunable topological interface states in a piezoelectric rod system with periodic electric boundary conditions. 
By adjusting the strain field, \citet{liu2018tunable} actively tuned the topological edge states in a 2D topological PC waveguides based on the quantum valley Hall effect.  \citet{feng2019magnetically} designed 1D magnetoelastic topological PC slabs and used a magnetic field to tune the topological interface states for Lamb waves contactlessly and nondestructively. 

Among the many studies on tunable topological PCs, soft topological PCs are receiving some attention because of their low operating frequency and the possibility of tunability by external loads.  \citet{li2018observation} presented topological interface states in a designed soft circular-hole PC plate that were dynamically tuned by altering the filling ratio and adjusting the external mechanical strain. \citet{huang2020flexible} showed that mechanical deformations can be used to actively tune the topological interface states in a 1D soft PC plate consisting of base elements with different topological properties. 
\citet{nguyen2019tunable} presented a 2D quantum valley Hall topological insulator, composed of soft cylinder inclusions and an elastic matrix. 
They found that the mechanical deformations can be exploited to modulate the topological properties of the structure and tune the topologically protected states. 
\citet{chen2021low} proposed a 1D soft waveguide composed of two topologically different PC elements, in which the low-frequency topological interface states for longitudinal waves were tuned by the axial force in a wide frequency range. 
Based on the quantum valley Hall effect, \citet{zhou2020voltage} designed a soft membrane-type PC consisting of a DE membrane and metallic particles, and broadened the frequency range of the topological interface mode in this voltage-controlled system.  

However, a high density of metallic particles can result in excessive deformations of the DE membrane, whose planar configuration may collapse. 
Motivated by the excellent electromechanical behaviors of DEs, here we design a 1D soft dielectric topological PC plate with step-wise cross-sections for the incremental bending waves, where the Bragg BGs are generated due to the geometric periodicity and the topological transition process can be realized by changing the geometric parameter. 
The low-frequency topological interface states in the soft dielectric PC plate can be actively tuned by applying electromechanical loads, which is a smarter and more convenient method to adjust the topological properties of PC waveguides compared with only applying a mechanical stimulus.

This paper is organized as follows. The basic formulations of nonlinear electroelasticity theory and its linearized incremental theory are summarized in Section~\ref{section2}. In Section~\ref{section3}, we analyze the nonlinear deformations of the designed soft dielectric PC plate with periodically varying cross-sections. By employing the Spectral Element Method, in Section~\ref{section4} we derive the dispersion relation and transmission coefficient for the incremental bending waves in the soft dielectric PC plate. The numerical results in Section~\ref{section5} show how the applied electric voltage and axial force affect the frequency of topological interface states in the soft dielectric plate waveguide. Some conclusions are made in Section~\ref{section6}.

\section{Preliminary Formulations}
\label{section2}

 This section briefly reviews the theoretical background of nonlinear electroelasticity and the related linearized incremental theory. For more detailed descriptions, interested readers are referred to the papers and textbook by \citet{dorfmann2006nonlinear, dorfmann2010electroelastic, dorfmann2014nonlinear}.


\subsection{Theory of nonlinear electroelasticity} \label{section2.1}


Consider an incompressible, soft deformable  electroelastic continuum. 
The \emph{undeformed reference} configuration is denoted by ${{\mathcal{B}}_{r}}$ with boundary $\partial {{\mathcal{B}}_{r}}$ and outward unit normal $\mathbf{N}$. 
A material particle in this configuration is labelled by its position vector $\mathbf{X}$. 
When subjected to an external load, the body deforms and occupies the \emph{deformed} \emph{current} configuration ${{\mathcal{B}}_{t}}$ at time $t$ with boundary $\partial {{\mathcal{B}}_{t}}$ and outward unit normal $\mathbf{n}_t$. 
The material point which was at $\mathbf{X}$ is now at $\mathbf{x}=\bm{\chi }\text{(}\mathbf{X},t\text{)}$, where $\bm{\chi }$ is an invertible vector function defined for all points in ${{\mathcal{B}}_{r}}$. The deformation gradient tensor is defined as $\mathbf{F}=\partial\mathbf{x}/\partial\mathbf{X} =\text{Grad}\bm{\chi }$ with its Cartesian components ${{{F}}_{i\alpha }}=\partial {{x}_{i}}/\partial {{X}_{\alpha }}$, where `Grad' is the gradient operator in ${{\mathcal{B}}_{r}}$.

In the absence of free charges and currents, and under the assumption of quasi-electrostatic approximation, the electric displacement vector $\mathbf{D}$ and electric field vector $\mathbf{E}$ in ${{\mathcal{B}}_{t}}$ satisfy Gauss's law and
Faraday's law, respectively, as
\begin{equation} \label{1}
{\rm{div}}{\kern 1pt} {\bf{D}} = 0,\qquad {\rm{   curl}}{\kern 1pt} {\bf{E}} = {\bf{0}},
\end{equation}
where `curl' and `div' are the curl and divergence operators defined in ${{\mathcal{B}}_{t}}$. In Eulerian form, the equation of motion without body force is given by
\begin{equation} \label{2}
{\rm{div}}{\kern 1pt} {\bm{\tau }} = \rho {{\bf{x}}_{,tt}},
\end{equation}
where $\mathbf{\bm{\tau }}$ is the `total Cauchy stress tensor' accounting for the contribution of electric body forces, $\rho$ is the material mass density, which remains constant due to the material incompressibility constraint $J=\det\mathbf{F}=1$, and the subscript $t$ following a comma signifies the material time derivative. The symmetry of $\mathbf{\bm{\tau }}$ is ensured by the conservation of angular momentum.

For an incompressible electroelastic material, the nonlinear constitutive relation can be derived from the total energy density function $\Omega (\mathbf{F},\bm{\mathcal{D}})$ per unit undeformed reference volume as \citep{dorfmann2006nonlinear}
\begin{equation} \label{3}
{\bf{T}} = \frac{{\partial \Omega }}{{\partial {\bf{F}}}}-{p_0}{{\bf{F}}^{-1}},\qquad \bm{{\cal E}} = \frac{{\partial \Omega }}{{\partial \bm{{\cal D}}}},
\end{equation}
where $\mathbf{T}={{\mathbf{F}}^{-1}}\mathbf{\bm{{\tau }}}$, $\mathcal{\bm{{\cal D}}}={{\mathbf{F}}^{-1}}\mathbf{D}$ and $\mathcal{\bm{{\cal E}}}={{\mathbf{F}}^{\text{T}}}\mathbf{E}$ are the `total' nominal stress tensor, Lagrangian electric displacement, and electric field vectors, respectively, and ${p_0}$ is a Lagrange multiplier accounting for the material incompressibility. 
Here, the superscript $^{\text{T}}$ signifies the transpose operation. 

For an incompressible isotropic electroelastic material, the energy density function can be written as $\Omega=\Omega(I_1,I_2,I_4,I_5,I_6)$, where  
\begin{equation} \label{4}
{I_1} = {\rm{tr}}{\bf{C}},\quad {I_2} = [{({\rm{tr}}{\bf{C}})^2} - {\rm{tr}}({{\bf{C}}^2})]/2,\quad {I_4} = {\bm{{\cal D}}} \cdot {\bm{{\cal D}}},\quad {I_5} = {\bm{{\cal D}}} \cdot ({\bf{C}}{\bm{{\cal D}}}),\quad {I_6} = {\bm{{\cal D}}} \cdot ({{\bf{C}}^2}{\bm{{\cal D}}}),
\end{equation}
and $\mathbf{C}={{\mathbf{F}}^{\text{T}}}\mathbf{F}$ is the right Cauchy-Green deformation tensor. Substitution of Eq.~\eqref{4} into Eq.~\eqref{3} leads to 
\begin{equation} \label{5}
\begin{array}{l}
{\bm{\tau }} = 2{\Omega _1}{\bf{b}} + 2{\Omega _2}({I_1}{\bf{b}} - {{\bf{b}}^2}) + 2{\Omega _5}{\bf{D}} \otimes {\bf{D}} + 2{\Omega _6}({\bf{D}} \otimes {\bf{bD}} + {\bf{bD}} \otimes {\bf{D}}) - {p_0}{\bf{I}},\\
{\bf{E}} = 2({\Omega _4}{{\bf{b}}^{ - 1}}{\bf{D}} + {\Omega _5}{\bf{D}} + {\Omega _6}{\bf{bD}}),
\end{array}
\end{equation}
where $\mathbf{b=F}{{\mathbf{F}}^{\text{T}}}$ is the left Cauchy-Green deformation tensor, $\mathbf{I}$ is the identity tensor, and ${{\Omega }_{m}}=\partial \Omega /\partial {{I}_{m}}  (m=1,2,4,5,6)$.

In this paper, we consider a dielectric elastomer plate coated by compliant electrodes on its surfaces, so that there is no electric field outside the material according to Gauss's theorem. 
As a result, the electric boundary conditions on $\partial {{\mathcal{B}}_{t}}$ are 
\begin{equation} \label{6}
{\bf{E}} \times {\bf{n}}_t = \bf{0},\qquad {\rm{   }}{\bf{D}} \cdot {\bf{n}}_t =  - {\sigma _{\rm{f}}},
\end{equation}
where ${{\sigma }_{\text{f}}}$ is the free charge density on $\partial {{\mathcal{B}}_{t}}$. In addition, the mechanical boundary condition on $\partial {{\mathcal{B}}_{t}}$ can be written in Eulerian form as
\begin{equation} \label{7}
{\mathbf{\bm{\tau n}}_t={{\mathbf{t}}^{a}}}, 
\end{equation}
where ${{\mathbf{t}}^{a}}$ is the applied mechanical traction vector per unit area of $\partial {{\mathcal{B}}_{t}}$.


\subsection{Linearized theory for incremental motions }\label{Sec2-2}


Following \citet{dorfmann2010electroelastic}, a time-dependent infinitesimal incremental motion $\dot{\mathbf{x}}\left( \mathbf{X},t \right)$, together with an incremental change $\dot {\bm{{\cal D}}}$ in the electric displacement vector is superimposed on an underlying static, finitely deformed configuration ${{\mathcal{B}}}$ with boundary $\partial {{\mathcal{B}}}$ and outward unit normal $\mathbf{n}$. 
Here and thereafter, a superposed dot is used to denote incremental quantities. 

The incremental governing equations can be written in the updated Lagrangian form as
\begin{equation} \label{8}
{\rm{div}}{\kern 1pt} {\dot {\bm{{\cal D}}}_0} = 0,\qquad {\rm{   curl}}{\kern 1pt}{\dot {\bm{{\cal E}}}_0} = \bf{0},\qquad {\rm{   div}}{{\bf{\dot T}}_0} = \rho {{\bf{u}}_{,tt}},
\end{equation}
where ${{\bf{\dot T}}_0} = {\bf{F\dot T}}$, 
${\dot {\bm{{\cal D}}}_0} = {\bf{F}}{\dot {\bm{{\cal D}}}}$ and  
${\dot {\bm{{\cal E}}}_0} = {{\bf{F}}^{ - {\rm{T}}}}{\dot {\bm{{\cal E}}}}$ are the `push-forward' versions of the Lagrangian increments  ${\bf{\dot T}}$, 
${\dot {\bm{{\cal D}}}}$ and 
${\dot {\bm{{\cal E}}}}$, respectively; $\mathbf{u}(\mathbf{x},t)=\dot{\mathbf{x}}\left( \mathbf{X},t \right)$ is the incremental mechanical displacement vector and the subscript 0 identifies the resulting `push-forward' quantities.

For an incompressible soft electroelastic material, the linearized incremental constitutive laws are
\begin{equation} \label{9}
{{\bf{\dot T}}_0} = {{\bm{{\cal A}}}_0}{\bf{H}} + {{\bm{{\cal M}}}_0}{\dot {\bm{{\cal D}}}_0} + {p_0}{\bf{H}} - {\dot p_0}{\bf{I}},\qquad {\rm{   }}{\dot {\bm{{\cal E}}}_0} =  {\bm{{\cal M}}}_0^{\rm{T}}{\bf{H}} + {{\bm{{\cal R}}}_0}{\dot {\bm{{\cal D}}}_0},
\end{equation}
where $\mathbf{H}=\text{grad}\mathbf{u}$ denotes the incremental displacement gradient tensor, `grad' is the gradient operator in ${{\mathcal{B}}}$, and ${\dot p_0}$ is an incremental change in the Lagrange multiplier $p_0$. 
In component form, the instantaneous electroelastic moduli tensors ${{\bm{{\cal A}}}_{0}}$, ${{\bm{{\cal M}}}_{0}}$ and ${{\bm{{\cal R}}}_{0}}$ are
\begin{align} \label{10}
& {{\cal A}_{0piqj}} ={F_{p\alpha }}{F_{q\beta }}{{\cal A}_{\alpha i\beta j}}={{\cal A}_{0qjpi}},\qquad {{\cal M}_{0piq}} = {F_{p\alpha }}F_{\beta q}^{ - 1}{{\cal M}_{\alpha i\beta }}={{\cal M}_{0ipq}}, \\
& {{\cal R}_{0ij}} = F_{\alpha i}^{ - 1}F_{\beta j}^{ - 1}{{\cal R}_{\alpha \beta }}={{\cal R}_{0ji}},
\end{align}
where the components of referential electroelastic moduli tensors $\bm{\mathcal{A}}$, $\bm{\mathcal{M}}$ and $\bm{\mathcal{R}}$ are defined as
\begin{equation} \label{11}
{{\cal A}_{\alpha i\beta j}} = \frac{{{\partial ^2}\Omega }}{{\partial {F_{i\alpha }}\partial {F_{j\beta }}}},\qquad {{\cal M}_{\alpha i\beta }} = \frac{{{\partial ^2}\Omega }}{{\partial {F_{i\alpha }}\partial {{\cal D}_\beta }}},\qquad {{\cal R}_{\alpha \beta }} = \frac{{{\partial ^2}\Omega }}{{\partial {{\cal D}_\alpha }\partial {{\cal D}_\beta }}}.
\end{equation}
Note that Greek and Roman indices are related to ${{\mathcal{B}}_{r}}$ and ${{\mathcal{B}}}$, respectively, and the summation convention for repeated indices is adopted. The incremental incompressibility condition requires 
\begin{equation} \label{12}
{\rm{div}}{\bf{u}} = {\rm{tr}}{\bf{H}} = 0.
\end{equation}

The incremental electric and mechanical boundary conditions, which are written in the updated Lagrangian form and are satisfied on $\partial {{\mathcal{B}}}$, are
\begin{equation} \label{13}
{\dot {\bm{{\cal E}}}_0} \times {\bf{n}} = \bf{0},\qquad {\dot {\bm{{\cal D}}}_0} \cdot {\bf{n}} =  - {\dot \sigma _{{\rm{F0}}}},\qquad {\rm{   }}{\bf{\dot T}}_0^{\rm{T}}{\bf{n}} = {\bf{\dot t}}_0^A,
\end{equation}
in which ${\dot \sigma _{{\rm{F0}}}}$ and ${\bf{\dot t}}_0^A$ indicate the incremental surface charge density and applied incremental mechanical traction per unit area of $\partial {{\mathcal{B}}}$, respectively, and the incremental electrical variables outside the electroelastic material are discarded.


\section{Nonlinear response of a homogeneous DE plate with step-wise cross-sections} \label{section3}

\begin{figure}[htbp]
	\centering	
	\includegraphics[width=0.8\textwidth]{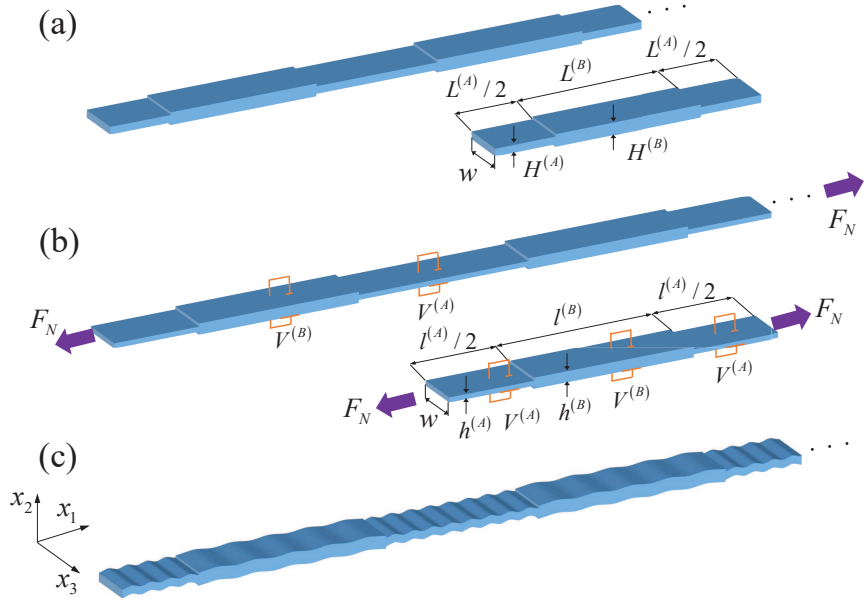}
	\caption{Sketch of an infinite soft dielectric PC plate with step-wise cross-sections: (a) undeformed configuration and its unit cell along with the geometric size; (b) deformed configuration and its unit cell along with the geometric size when subjected to an axial force ${{F}_{N}}$, electric voltages ${V ^{\left( A \right)}}$ and ${V ^{\left( B \right)}}$ in the thickness direction; (c) incremental bending waves superimposed on the static finite deformations shown in (b).}
	\label{Fig1}
\end{figure}

A \emph{single-phase} thin soft dielectric PC plate with periodically varying, step-wise square cross-sections is shown in Fig.~\ref{Fig1}. 
In the undeformed configuration (Fig.~\ref{Fig1}(a)), the unit cell composed of a homogeneous incompressible DE has two thinner sub-plates $A$ and $C$ with equal length ${L^{\left( A \right)}}/2$ and height  ${H^{\left( A \right)}}$, sandwiching a thicker sub-plate $B$ of length ${L^{\left( B \right)}}$ and height ${H^{\left( B \right)}}$ (we assume ${H^{\left( B \right)}} > {H^{\left( A \right)}}$). The physical quantities of the sub-plate $p$ ($p=A,B,C$) are denoted by the superscript ${{\left( \cdot  \right)}^{\left( p \right)}}$ throughout this paper. Geometrically, the total length of undeformed unit cell is $L = {L^{\left( A \right)}} + {L^{\left( B \right)}}$ along the ${x_1}$ direction, and different sub-plates have an equal width $w = {w^{\left( A \right)}} = {w^{\left( B \right)}} = {w^{\left( C \right)}}$. In addition, the unit cell of the PC plate is designed to be symmetric with respect to the neutral ${x_1}$-${x_3}$ plane in order to decouple the longitudinal and bending (transverse) waves \citep{yin2018band}. The difference in cross-sectional areas helps construct the geometric periodicity and results in the Bragg BGs for this periodic soft dielectric PC plate.

When subjected to electric voltages applied between the compliant electrodes on its top and bottom surfaces, combined with an axial force $F_N$, the soft dielectric PC plate is activated to the deformed configuration, as shown in Fig.~\ref{Fig1}(b). The application of electric voltage across the thickness of the sub-plates generates electrostatic forces, which decrease the thickness and increase the in-plane (${x_1}$-${x_3}$ plane) size. The tensile axial force has a similar influence on the geometry of the soft dielectric PC plate. Here, the following assumptions are introduced: (1) For the external loads keeping unchanged in the width direction, we assume the plane-strain state for the thin dielectric plate with a large width (compared with the height); (2) The nonlinear deformations are approximately assumed uniform in each sub-plate of the unit cell, because inhomogeneous local deformations are confined to small regions near the interfaces between different sub-plates and hardly affect the response of low-frequency topological interface states. The latter assumption has been validated by the finite element simulations for longitudinal waves in soft PC cylinders \citep{chen2021low}.

In the Cartesian coordinate system with coordinates $x_1$, $x_2$ and $x_3$ along the length, thickness, and width directions of the plate, respectively, the plane-strain nonlinear deformation of each sub-plate can be described by the uniform deformation gradient tensor as
\begin{equation} \label{14}
{{\bf{F}}^{\left( p \right)}} = {\rm{diag}}[{\lambda ^{\left( p \right)}},1/{\lambda ^{\left( p \right)}},1],
\end{equation}
where ${\lambda ^{\left( p \right)}}$ is the principal stretch of sub-plate $p$ in the $x_1$ direction. Accordingly, the  geometric sizes of each sub-plate in the deformed unit cell are
\begin{equation} \label{15}
{l^{\left( p \right)}} = {\lambda ^{\left( p \right)}}{L^{\left( p \right)}},\qquad {h^{\left( p \right)}} = {H^{\left( p \right)}}/{\lambda ^{\left( p \right)}},
\end{equation}
where ${h ^{\left( p \right)}}$, $l^{\left( A \right)}/2=l^{\left( C \right)}/2$, and ${l ^{\left( B \right)}}$ are the thickness of sub-plate $p$ and the lengths of the thinner and thicker sub-plates in the deformed state, respectively. 
Note that we have assumed that sub-plates $A$ and $C$ are subjected to the same voltage, leading to $l^{\left( A \right)}/2=l^{\left( C \right)}/2$. Consequently, the length of the deformed unit cell is $l=l^{\left( A \right)}+l^{\left( B \right)}$. 
Because the electric voltage ${V ^{\left( p \right)}}$ is applied along the $x_2$ direction, the Eulerian electric displacement vector $\mathbf{D}$ has only one nonzero component, $D_2^{\left( p \right)}$.

To investigate bending wave propagation, we consider that the soft dielectric PC plate is modelled by the  incompressible Gent ideal dielectric model \citep{gent1996new, wu2018tuning},
\begin{equation} \label{16}
{\Omega ^{\left( p \right)}} =  - \frac{{\mu {J_m}}}{2}\ln \left( {1 - \frac{{I_1^{\left( p \right)} - 3}}{{{J_m}}}} \right) + \frac{{I_5^{\left( p \right)}}}{{2\varepsilon }},
\end{equation}
where $\mu$ and $\varepsilon  = {\varepsilon _0}{\varepsilon _r}$ are the shear modulus and material permittivity in the undeformed state, respectively, with the vacuum permittivity $\varepsilon _0=8.85\text{ pF/m}$ and $\varepsilon _r$ being the relative permittivity, and $J_m$ is a dimensionless constant characterizing the strain-stiffening effect of the DE plate. 
The neo-Hookean ideal dielectric model can be recovered from Eq.~\eqref{16} in the limit of ${J_m} \to \infty $. 
Here, $I_1^{\left( p \right)} = {\left( {{\lambda ^{\left( p \right)}}} \right)^2} + {\left( {{\lambda ^{\left( p \right)}}} \right)^{ - 2}}{\rm{ + 1}}$ and 
$I_5^{\left( p \right)} = {\left( {D_2^{\left( p \right)}} \right)^2}$ are the relevant invariants. 
For the \emph{single phase} soft dielectric PC plate, the electromechanical material parameters $\mu$, $J_m$ and $\varepsilon$ are the same for three sub-plates, but the stretch ${\lambda ^{\left( p \right)}}$ could be different due to the non-uniform cross-section.

By substituting Eq.~\eqref{16} into Eq.~\eqref{5}, the expressions for Cauchy stress components $\tau _{11}^{\left( p \right)}$ and $\tau _{22}^{\left( p \right)}$ as well as for the Eulerian electric field component in the $x_2$ direction are obtained as
\begin{equation} \label{17}
\begin{split}
\tau _{11}^{\left( p \right)} &= \frac{{\mu {J_m}}}{{{J_m} - I_1^{\left( p \right)} + 3}}{\left( {{\lambda ^{\left( p \right)}}} \right)^2} - p_0^{\left( p \right)},\\
\tau _{22}^{\left( p \right)} &= \frac{{\mu {J_m}}}{{{J_m} - I_1^{\left( p \right)} + 3}}{\left( {{\lambda ^{\left( p \right)}}} \right)^{ - 2}}{\rm{ + }}\frac{{{{\left( {D_2^{\left( p \right)}} \right)}^2}}}{\varepsilon } - p_0^{\left( p \right)}, \\
E_2^{\left( p \right)} &= \frac{{D_2^{\left( p \right)}}}{\varepsilon }.
\end{split}
\end{equation}

To balance the axial force $F_N$ and fulfil the traction-free conditions on the top and bottom surfaces (at ${x_2} =  \pm {h^{\left( p \right)}}/2$ with origin located at the centroid of the plate), the stress components $\tau _{11}^{\left( p \right)}$ and $\tau _{22}^{\left( p \right)}$ must satisfy
\begin{equation} \label{18}
\tau _{11}^{\left( p \right)}=
\frac{{{F_N}}}{{a^{\left( p \right)}}},\qquad \tau _{22}^{\left( p \right)}=0,
\end{equation}
where $a^{\left( p \right)}=w{h^{\left( p \right)}}$ is the cross-sectional area of sub-plate $p$ in the deformed configuration. 
Furthermore, because the voltage ${V ^{\left( p \right)}}$ is applied across the thickness, the only nonzero component of the Eulerian electric field vector is
\begin{equation} \label{19}
E_2^{\left( p \right)} = \frac{{{V^{\left( p \right)}}}}{{{h^{\left( p \right)}}}},
\end{equation}
showing that  $E_2^{\left( p \right)}$ is linearly related to the voltage $V^{\left( p \right)}$.

Inserting Eqs.~\eqref{18}$_2$ and \eqref{19} into Eq.~\eqref{17}$_{2,3}$ yields the Lagrange multiplier $p_0^{\left( p \right)}$ of each sub-plate as
\begin{equation} \label{20-1}
p_0^{\left( p \right)} = \frac{{\mu {J_m}}}{{{J_m} - I_1^{\left( p \right)} + 3}} {{\left( {{\lambda ^{\left( p \right)}}} \right)}^{ - 2}}  + \varepsilon {\left( {\frac{{{V^{\left( p \right)}}}}{{{h^{\left( p \right)}}}}} \right)^2}.
\end{equation}
Substituting Eqs.~\eqref{17}$_1$ and \eqref{20-1} into Eq.~\eqref{18}$_1$, we then obtain the nonlinear relation between 
${\lambda ^{\left( p \right)}}$, ${V ^{\left( p \right)}}$ and $F_N$ for sub-plate $p$ as
\begin{equation} \label{20}
\frac{{\mu {J_m}}}{{{J_m} - I_1^{\left( p \right)} + 3}}\left[ {{{\left( {{\lambda ^{\left( p \right)}}} \right)}^2} - {{\left( {{\lambda ^{\left( p \right)}}} \right)}^{ - 2}}} \right] - \varepsilon {\left( {\frac{{{V^{\left( p \right)}}}}{{{h^{\left( p \right)}}}}} \right)^2} = \frac{{{F_N}}}{{{w{h^{\left( p \right)}}}}}.
\end{equation}
According to Eq.~\eqref{20}, the principal stretch ratio ${\lambda ^{\left( p \right)}}$ of each sub-plate can be found once the external loads ${V ^{\left( p \right)}}$ and $F_N$ are prescribed. Usually, the resulting stretch ratios for different sub-plates $A$ and $B$ are not the same. 


\section{Incremental bending waves in a soft dielectric PC plate} 
\label{section4}


Based on the nonlinear deformations obtained in Sec.~\ref{section3}, we now study the superimposed incremental bending wave (see Fig.~\ref{Fig1}(c)), propagating in the dielectric PC plate subjected to the voltage and axial force. 
We derive the incremental equation of motion and its solutions in Sec.~\ref{Sec4-1}. In Sec.~\ref{Sec4-2} and Sec.~\ref{Sec4-3}, we use the Spectral Element Method \citep{lee2009spectral, han2012modified} to derive in turn the dispersion relation of an infinite soft PC plate and the transmission coefficient of a finite soft PC plate waveguide.


\subsection{Incremental motion equation and its solutions} 
\label{Sec4-1}


In this subsection, we drop the superscript notation ${{\left( \cdot  \right)}^{\left( p \right)}}$ of each sub-plate  temporarily for compactness. 
Following \citet{gei2010controlling} and \citet{shmuel2016manipulating}, we assume that the perturbations in the electric quantities generated by the incremental wave motions are negligible. 
We assume that the plane-strain state holds (all physical quantities are independent of coordinate $x_3$) and that the deflection $u_2$ depends on coordinate $x_1$ only. 
Then the equation of motion for bending waves in the soft dielectric PC plate under the axial force is analogous to that found for a pre-stressed Euler-Bernoulli beam. 

Neglecting the incremental electric quantities, the components ${{\dot T}_{011}}$ and ${{\dot T}_{022}}$ can be obtained by the incremental constitutive relation \eqref{9}$_1$ as 
\begin{equation} \label{21}
\begin{array}{l}
{{\dot T}_{011}} = \left( {{{\cal A}_{01111}}{\rm{ + }}{p_0}} \right){H_{11}} + {{\cal A}_{01122}}{H_{22}} - {{\dot p}_0}, \qquad {{\dot T}_{022}} = {{\cal A}_{01122}}{H_{11}} + \left( {{{\cal A}_{02222}}{\rm{ + }}{p_0}} \right){H_{22}} - {{\dot p}_0},
\end{array}
\end{equation}
where we  used ${H_{33}}=0$ (plane-strain assumption), and the components of ${{\bm{{\cal A}}}_{0}}$ for the Gent ideal dielectric model characterized by Eq.~\eqref{16} can be derived from Eq.~\eqref{11}$_1$ as
\begin{equation} \label{22}
\begin{array}{l}
{{\cal A}_{{\rm{01111}}}} = \dfrac{{2\mu {J_m}}}{{{{\left( {{J_m} - {I_{\rm{1}}} + 3} \right)}^2}}}{\lambda ^4} + \dfrac{{\mu {J_m}}}{{{J_m} - {I_{\rm{1}}} + 3}}{\lambda ^2},\qquad {{\cal A}_{01122}} = \dfrac{{2\mu {J_m}}}{{{{\left( {{J_m} - {I_1} + 3} \right)}^2}}}, \vspace{1ex} \\
{{\cal A}_{{\rm{02222}}}} = \dfrac{{2\mu {J_m}}}{{{{\left( {{J_m} - {I_{\rm{1}}} + 3} \right)}^2}}}{\lambda ^{ - 4}} + \dfrac{{\mu {J_m}}}{{{J_m} - {I_1} + 3}}{\lambda ^{ - 2}} + \dfrac{1}{\varepsilon }{\left( {{D_2}} \right)^2}.
\end{array}
\end{equation}
Making use of the incremental incompressibility constraint ${H_{11}} + {H_{22}} = 0$ from Eq.~\eqref{12}, Eq.~\eqref{21} is expressed in terms of ${H_{11}}$ and ${{\dot p}_0}$ as 
\begin{equation} \label{23}
\begin{array}{l}
{{\dot T}_{011}} = \left( {{{\cal A}_{01111}} + {p_0} - {{\cal A}_{01122}}} \right){H_{11}} - {{\dot p}_0}, \qquad {{\dot T}_{022}} = \left( {{{\cal A}_{01122}} - {{\cal A}_{02222}} - {p_0}} \right){H_{11}} - {{\dot p}_0}.
\end{array}
\end{equation}

For bending waves propagating in the Euler-Bernoulli beam or plate along the $x_1$ direction, the total bending moment ${\bf{m}}_t$ is  
\begin{equation} \label{24}
\begin{split}
{\bf{m}}_t &= \int_{\partial {{\cal B}_r}} {\bf{x}}  \times {\left( {{\bf{T}} + {\bf{\dot T}}} \right)^{\rm{T}}}{\bf{N}}{\rm{d}}A = \int_{\partial {{\cal B}}} {\bf{x}}  \times {\left( {{\bf{T}} + {\bf{\dot T}}} \right)^{\rm{T}}}{{\bf{F}}^{\rm{T}}}{\bf{n}}{\rm{d}}a\\
&= \int_{\partial {{\cal B}}} {{x_2}{{\bf{i}}_2}}  \times {\left( {{\bf{T}} + {\bf{\dot T}}} \right)^{\rm{T}}}{{\bf{F}}^{\rm{T}}}{\bf{n}}{\rm{d}}a,
\end{split}
\end{equation}
in which ${{\bf{i}}_1}$, ${{\bf{i}}_2}$ and ${{\bf{i}}_3}$ are the unit vectors along the $x_1$, $x_2$ and $x_3$ directions, respectively, ${\partial {{\cal B}}}$ is the cross-section of the deformed plate with its normal being $\bf{n}={{\bf{i}}_1}$, and we used Nanson's formula $\mathbf{n} \text{d}a=J{{\bf{F}}^{ - {\rm{T}}}}{\bf{N}}{\rm{d}}A={{\bf{F}}^{ - {\rm{T}}}}{\bf{N}}{\rm{d}}A$, where $\text{d}a$ and $\text{d}A$ are infinitesimal surface elements in ${{\mathcal{B}}}$ and ${{\mathcal{B}_r}}$, respectively. 
Because the nominal stress component ${T_{11}}$ (or Cauchy stress ${\sigma _{11}}$) is a constant, the initial bending moment ${\bf{m}}_i$ satisfies
\begin{equation} \label{25}
{\bf{m}}_i = \int_{\partial {\cal B}} {{x_2}{{\bf{i}}_2}}  \times {{\bf{T}}^{\rm{T}}}{{\bf{F}}^{\rm{T}}}{\bf{n}}{\rm{d}}a =  - w\int_{ - h/2}^{h/2} {{T_{11}}\lambda {x_2}{\rm{d}}{x_2}} {{\bf{i}}_3} = \bf{0}.
\end{equation}
Inserting Eq.~\eqref{25} into Eq.~\eqref{24} yields the incremental bending moment ${\bf{m}}={\bf{m}}_t-{\bf{m}}_i$ as
\begin{equation} \label{26}
\begin{split}
{\bf{m}} &= \int_{\partial {\cal B}} {{x_2}{{\bf{i}}_2}}  \times {{{\bf{\dot T}}}^{\rm{T}}}{{\bf{F}}^{\rm{T}}}{\bf{n}}{\rm{d}}a = \int_{\partial {\cal B}} {{x_2}{{\bf{i}}_2}}  \times {\bf{\dot T}}_0^{\rm{T}}{\bf{n}}{\rm{d}}a =  - w\int_{ - h/2}^{h/2} {{{\dot T}_{011}}{x_2}{\rm{d}}{x_2}} {{\bf{i}}_3}.
\end{split}
\end{equation}

It is well-accepted that the shear deformation can be ignored in the Euler-Bernoulli beam theory, which gives  
\begin{equation} \label{27}
{H_{12}} + {H_{21}} =  {u_{1,2}} + {u_{2,1}} = 0.
\end{equation}
Considering that the plate is very thin, we can assume that the deflection $u_2$ keeps unchanged along the thickness direction (independent of $x_2$). 
The differentiation of Eq.~\eqref{27} with respect to $x_1$ along with a further integration with respect to  $x_2$ leads to
\begin{equation} \label{28}
{H_{11}} = {u_{1,1}} =  - {x_2}{u_{2,11}} + {C_0},
\end{equation}
where $u_{2,11}$ is assumed to be uniform along $x_2$ and $C_0$ is an undetermined constant. Substituting Eqs.~\eqref{23}$_1$ and \eqref{28} into Eq.~\eqref{26}, we obtain the expression of the incremental bending moment as
\begin{equation} \label{29}
{\bf{m}} = {u_{2,11}}\left( {{{\cal A}_{01111}} + {p_0} - {{\cal A}_{01122}}} \right)I{{\bf{i}}_3} + w\int_{ - h/2}^{h/2} {{{\dot p}_0}{x_2}{\rm{d}}{x_2}} {{\bf{i}}_3},
\end{equation}
where 
$I = w\int_{ - h/2}^{h/2} {x_2^2{\rm{d}}{x_2}}  = {w{h^3}}/{{12}}$ is the second moment of area. 
The second term in Eq.~\eqref{29} depends on 
${\dot p_0}$, which is  determined by the incremental boundary condition, as follows. 
The soft PC plate is free of mechanical traction at the top and bottom surfaces, and  ${\dot T_{022}} = 0 $ there. 
Therefore, ${\dot p_0}$ is derived from Eq.~\eqref{23}$_2$ as 
\begin{equation} \label{30}
{{\dot p}_0} = \left( {{{\cal A}_{01122}} - {{\cal A}_{02222}} - {p_0}} \right){H_{11}}.
\end{equation}
By inserting Eq.~\eqref{30} into Eq.~\eqref{29}, we obtain the relation between the incremental bending moment ${\bf{m}}$ and the deflection $u_2$ as
\begin{equation} \label{31}
{\bf{m}} = {u_{2,11}}{\cal A}_0^{\rm{e}}I{{\bf{i}}_3},
\end{equation}
where ${\cal A}_0^{\rm{e}} = {{\cal A}_{01111}} + {{\cal A}_{02222}} + 2{p_0} - 2{{\cal A}_{01122}}$, and the Lagrange multiplier $p_0$ is determined by Eq.~\eqref{20-1}.

Similarly to the derivations of governing equations for transverse waves in a pre-stressed Euler-Bernoulli beam \citep{graff2012wave}, the balance of linear momentum along with Eq.~\eqref{31} is applied to a differential element of the soft dielectric plate. As a result, the incremental bending wave equation of the soft dielectric plate takes the form
\begin{equation} \label{32}
- {\cal A}_0^{\rm{e}}I{u_{2,1111}} + {F_N}{u_{2,11}} = \rho a{u_{2,tt}},
\end{equation}
where $a$ is the cross-sectional area of the deformed plate. For the superimposed bending wave motions, the deflection $u_2$ is only dependent on $x_1$ and $t$, and the general solution to Eq.~\eqref{32} is
\begin{equation} \label{33}
{u_2} = \left( {{W_1}{{\rm{e}}^{{\rm{i}}\alpha {x_1}}} + {W_2}{{\rm{e}}^{ - {\rm{i}}\alpha {x_1}}} + {W_3}{{\rm{e}}^{{\rm{i}}\beta {x_1}}} + {W_4}{{\rm{e}}^{ - {\rm{i}}\beta {x_1}}}} \right){{\rm{e}}^{ - {\rm{i}}\omega t}},
\end{equation}
where $\rm{i}$ is the imaginary unit, $\omega$ is the angular frequency, $W_1$, $W_2$, $W_3$ and $W_4$ are the arbitrary amplitude constants, and
\begin{equation} \label{34}
\alpha  = \sqrt {\dfrac{{ - {F_N} + \sqrt {{F_N}^2 + 4{\cal A}_0^{\rm{e}}I\rho a{\omega ^2}} }}{{2{\cal A}_0^{\rm{e}}I}}} , \qquad 
\beta  = {\rm{i}}\sqrt {\dfrac{{{F_N} + \sqrt {{F_N}^2 + 4{\cal A}_0^{\rm{e}}I\rho a{\omega ^2}} }}{{2{\cal A}_0^{\rm{e}}I}}}.
\end{equation}
In general, ${\cal A}_0^{\rm{e}}$, $I$, $a$, $\alpha$ and $\beta$ are step-wise constants for the soft dielectric PC plate which depend on the external electromechanical loads ${V}$ and $F_N$, and they are usually not the same for different sub-plates. 


\subsection{Dispersion relation of an infinite soft PC plate} \label{Sec4-2}


Now we shed light on the derivation of the dispersion relation for an infinite soft PC plate with step-wise cross-sections. To describe the mechanical state at each section of the PC plate, the deflection, rotation angle, bending moment and shear force at that section are introduced and expressed in the following forms (the harmonic time-dependency ${{\rm{e}}^{ - {\rm{i}}\omega t}}$ is assumed for all fields and is omitted for simplicity):
\begin{equation} \label{35}
\begin{split}
u_2^{\left( p \right)}\left( {{x^{(p)}_1}} \right) &= W_1^{\left( p \right)}{{\rm{e}}^{{\rm{i}}{\alpha ^{\left( p \right)}}{x^{(p)}_1}}} + W_2^{\left( p \right)}{{\rm{e}}^{ - {\rm{i}}{\alpha ^{\left( p \right)}}{x^{(p)}_1}}} + W_3^{\left( p \right)}{{\rm{e}}^{{\rm{i}}{\beta ^{\left( p \right)}}{x^{(p)}_1}}} + W_4^{\left( p \right)}{{\rm{e}}^{ - {\rm{i}}{\beta ^{\left( p \right)}}{x^{(p)}_1}}},\\
{\varphi ^{\left( p \right)}}\left( {{x^{(p)}_1}} \right) &= u_{2,1}^{\left( p \right)},\\
{{\rm{m}}^{\left( p \right)}}\left( {{x^{(p)}_1}} \right) &= 
{\zeta ^{\left( p \right)}}u_{2,11}^{\left( p \right)},\\
{Q^{\left( p \right)}}\left( {{x^{(p)}_1}} \right) &=  - {\zeta ^{\left( p \right)}}u_{2,111}^{\left( p \right)},
\end{split}
\end{equation}
respectively, where ${\zeta ^{\left( p \right)}} = {\cal A}_0^{{\rm{e}}\left( p \right)}{I^{\left( p \right)}}$, and $x^{(p)}_1$ is the local coordinate of each sub-plate. 

To avoid the severe numerical instabilities of the traditional Transfer-Matrix Method (refer to \citet{rokhlin2002stable} and \citet{pao2007reverberation} for more details on the rigorous analysis of numerical instabilities) when studying the bending/transverse wave propagation characteristics (especially the transmission spectrum) of soft PC plates, we employ the stable Spectral Element Method \citep{lee2009spectral, han2012modified} to derive the bending wave propagation behaviors of soft PC plates. It is worth noting that the Spectral Element Method adopted in this paper differs from that used by \citet{wang2021manipulation,wang2021precise}, where they developed the so-called Spectral Element Method with exponential convergence, by adopting interpolation nodes within the Finite Element Method framework.
Our method is also different from the stable Hybrid Compliance-Stiffness Matrix Method used by \citet{shmuel2016manipulating} for flexural waves in two-component soft dielectric films.

We first introduce the nodal displacement vector 
${{\bf{q}}^{\left( p \right)}}= [{{\bf{q}}_L^{\left( p \right){\rm{T}}}} \quad
{{\bf{q}}_R^{\left( p \right){\rm{T}}}}]^\textrm{T}$ of sub-plate $p$ with ${\bf{q}}_L^{\left( p \right)}=[{u_2^{\left( p \right)}\left( 0 \right)} \quad {{\varphi ^{\left( p \right)}}\left( 0 \right)}]^\textrm{T} $ and ${\bf{q}}_R^{\left( p \right)}= [{u_2^{\left( p \right)}\left( {{{\tilde l}^{\left( p \right)}}} \right)} \quad {{\varphi ^{\left( p \right)}}\left( {{{\tilde l}^{\left( p \right)}}} \right)}]^\textrm{T}$, where the subscript $L$ and $R$ stand for the physical quantities at the left and right ends of each sub-plate, and ${{\tilde l}^{\left( p \right)}}={l^{\left( A \right)}}/2$ (or ${l^{\left( B \right)}}$) is associated with sub-plates A and C (or sub-plate B). According to Eq.~\eqref{35}$_{1,2}$, the nodal displacement vector 
${{\bf{q}}^{\left( p \right)}}$ can be written as 
\begin{equation} \label{36}
{{\bf{q}}^{\left( p \right)}} = {{\bf{U}}^{\left( p \right)}}{{\bf{c}}^{\left( p \right)}},
\end{equation}
where the $4\times4$ matrix ${{\bf{U}}^{\left( p \right)}}$ and the coefficient vector ${{\bf{c}}^{\left( p \right)}}$ are
\begin{equation} \label{37}
\begin{split}
{{\bf{U}}^{\left( p \right)}} &= \left[ {\begin{array}{*{20}{c}}
	1&1&1&1\\
	{{\rm{i}}{\alpha ^{\left( p \right)}}}&{ - {\rm{i}}{\alpha ^{\left( p \right)}}}&{{\rm{i}}{\beta ^{\left( p \right)}}}&{ - {\rm{i}}{\beta ^{\left( p \right)}}}\\
	{{{\rm{e}}^{{\rm{i}}{\alpha ^{\left( p \right)}}{{\tilde l}^{\left( p \right)}}}}}&{{{\rm{e}}^{ - {\rm{i}}{\alpha ^{\left( p \right)}}{{\tilde l}^{\left( p \right)}}}}}&{{{\rm{e}}^{{\rm{i}}{\beta ^{\left( p \right)}}{{\tilde l}^{\left( p \right)}}}}}&{{{\rm{e}}^{ - {\rm{i}}{\beta ^{\left( p \right)}}{{\tilde l}^{\left( p \right)}}}}}\\
	{{\rm{i}}{\alpha ^{\left( p \right)}}{{\rm{e}}^{{\rm{i}}{\alpha ^{\left( p \right)}}{{\tilde l}^{\left( p \right)}}}}}&{ - {\rm{i}}{\alpha ^{\left( p \right)}}{{\rm{e}}^{ - {\rm{i}}{\alpha ^{\left( p \right)}}{{\tilde l}^{\left( p \right)}}}}}&{{\rm{i}}{\beta ^{\left( p \right)}}{{\rm{e}}^{{\rm{i}}{\beta ^{\left( p \right)}}{{\tilde l}^{\left( p \right)}}}}}&{ - {\rm{i}}{\beta ^{\left( p \right)}}{{\rm{e}}^{ - {\rm{i}}{\beta ^{\left( p \right)}}{{\tilde l}^{\left( p \right)}}}}}
	\end{array}} \right],\\
{{\bf{c}}^{\left( p \right)}} &= {\left[ {\begin{array}{*{20}{c}}
		{W_1^{\left( p \right)}}&{W_2^{\left( p \right)}}&{W_3^{\left( p \right)}}&{W_4^{\left( p \right)}}
		\end{array}} \right]^{\rm{T}}},
\end{split}
\end{equation}
respectively.
 Similarly, the nodal force vector ${{\bf{f}}^{\left( p \right)}}= [{{\bf{f}}_L^{\left( p \right){\rm{T}}}} \quad
{{\bf{f}}_R^{\left( p \right){\rm{T}}}}]^\textrm{T}$ can be expressed as
\begin{equation} \label{38}
{{\bf{f}}^{\left( p \right)}} = {{\bf{V}}^{\left( p \right)}}{{\bf{c}}^{\left( p \right)}},
\end{equation}
where ${\bf{f}}_L^{\left( p \right)}= [{ - {{\rm{m}}^{\left( p \right)}}\left( 0 \right)} \quad { - {Q^{\left( p \right)}}\left( 0 \right)}]^\textrm{T}$, ${\bf{f}}_R^{\left( p \right)}= [ 
{{{\rm{m}}^{\left( p \right)}}\left( {{{\tilde l}^{\left( p \right)}}} \right)} \quad 
{{Q^{\left( p \right)}}\left( {{{\tilde l}^{\left( p \right)}}} \right)}]^\textrm{T}$, and 
\begin{small}
\begin{equation} \label{39}
\begin{array}{l}
{{\bf{V}}^{\left( p \right)}} = \\[4pt]
 \left[ {\begin{array}{*{20}{c}}
	{{\zeta ^{\left( p \right)}}{{\left( {{\alpha ^{\left( p \right)}}} \right)}^2}}&{{\zeta ^{\left( p \right)}}{{\left( {{\alpha ^{\left( p \right)}}} \right)}^2}}&{{\zeta ^{\left( p \right)}}{{\left( {{\beta ^{\left( p \right)}}} \right)}^2}}&{{\zeta ^{\left( p \right)}}{{\left( {{\beta ^{\left( p \right)}}} \right)}^2}}\\
	{ - {\rm{i}}{\zeta ^{\left( p \right)}}{{\left( {{\alpha ^{\left( p \right)}}} \right)}^3}}&{{\rm{i}}{\zeta ^{\left( p \right)}}{{\left( {{\alpha ^{\left( p \right)}}} \right)}^3}}&{ - {\rm{i}}{\zeta ^{\left( p \right)}}{{\left( {{\beta ^{\left( p \right)}}} \right)}^3}}&{{\rm{i}}{\zeta ^{\left( p \right)}}{{\left( {{\beta ^{\left( p \right)}}} \right)}^3}}\\
	{ - {\zeta ^{\left( p \right)}}{{\left( {{\alpha ^{\left( p \right)}}} \right)}^2}{{\rm{e}}^{{\rm{i}}{\alpha ^{\left( p \right)}}{{\tilde l}^{\left( p \right)}}}}}&{ - {\zeta ^{\left( p \right)}}{{\left( {{\alpha ^{\left( p \right)}}} \right)}^2}{{\rm{e}}^{ - {\rm{i}}{\alpha ^{\left( p \right)}}{{\tilde l}^{\left( p \right)}}}}}&{ - {\zeta ^{\left( p \right)}}{{\left( {{\beta ^{\left( p \right)}}} \right)}^2}{{\rm{e}}^{{\rm{i}}{\beta ^{\left( p \right)}}{{\tilde l}^{\left( p \right)}}}}}&{ - {\zeta ^{\left( p \right)}}{{\left( {{\beta ^{\left( p \right)}}} \right)}^2}{{\rm{e}}^{ - {\rm{i}}{\beta ^{\left( p \right)}}{{\tilde l}^{\left( p \right)}}}}}\\
	{{\rm{i}}{\zeta ^{\left( p \right)}}{{\left( {{\alpha ^{\left( p \right)}}} \right)}^3}{{\rm{e}}^{{\rm{i}}{\alpha ^{\left( p \right)}}{{\tilde l}^{\left( p \right)}}}}}&{ - {\rm{i}}{\zeta ^{\left( p \right)}}{{\left( {{\alpha ^{\left( p \right)}}} \right)}^3}{{\rm{e}}^{ - {\rm{i}}{\alpha ^{\left( p \right)}}{{\tilde l}^{\left( p \right)}}}}}&{{\rm{i}}{\zeta ^{\left( p \right)}}{{\left( {{\beta ^{\left( p \right)}}} \right)}^3}{{\rm{e}}^{{\rm{i}}{\beta ^{\left( p \right)}}{{\tilde l}^{\left( p \right)}}}}}&{ - {\rm{i}}{\zeta ^{\left( p \right)}}{{\left( {{\beta ^{\left( p \right)}}} \right)}^3}{{\rm{e}}^{ - {\rm{i}}{\beta ^{\left( p \right)}}{{\tilde l}^{\left( p \right)}}}}}
	\end{array}} \right].
\end{array}
\end{equation}
\end{small}
Combining Eq.~\eqref{36} with Eq.~\eqref{38}, we obtain the following relation:
\begin{equation} \label{40}
{{\bf{f}}^{\left( p \right)}} = {{\bf{G}}^{\left( p \right)}}{{\bf{q}}^{\left( p \right)}} = \left[ {\begin{array}{*{20}{c}}
	{{\bf{G}}_{LL}^{\left( p \right)}}&{{\bf{G}}_{LR}^{\left( p \right)}} \vspace{1ex}\\
	{{\bf{G}}_{RL}^{\left( p \right)}}&{{\bf{G}}_{RR}^{\left( p \right)}}
	\end{array}} \right]{{\bf{q}}^{\left( p \right)}},
\end{equation}
where 
${{\bf{G}}^{\left( p \right)}} = {{\bf{V}}^{\left( p \right)}}{\left( {{{\bf{U}}^{\left( p \right)}}} \right)^{ - 1}}$ is the $4\times4$ dynamic stiffness matrix for sub-plate $p$ ($p=A, B, C$).

For each deformed unit cell, the incremental continuity conditions at the interfaces between different sub-plates can be expressed as
\begin{equation} \label{41}
\begin{array}{l}
{\bf{f}}_R^{\left( A \right)} =  - {\bf{f}}_L^{\left( B \right)},\quad {\bf{q}}_R^{\left( A \right)} = {\bf{q}}_L^{\left( B \right)}, \vspace{1ex}\\
{\bf{f}}_R^{\left( B \right)} =  - {\bf{f}}_L^{\left( C \right)},\quad {\bf{q}}_R^{\left( B \right)} = {\bf{q}}_L^{\left( C \right)}.
\end{array}
\end{equation}
Substituting Eq.~\eqref{41} into Eq.~\eqref{40}, we derive the relation between the nodal force vector and nodal displacement vector at the left and right ends of each unit cell, which we write in matrix form as 
\begin{equation} \label{42}
\left[ {\begin{array}{*{20}{c}}
	{{\bf{f}}_L^{\left( A \right)}}\vspace{1ex}\\
	{{\bf{f}}_R^{\left( C \right)}}
	\end{array}} \right] = \left[ {\begin{array}{*{20}{c}}
	{{{\bf{K}}_{LL}}}&{{{\bf{K}}_{LR}}}\vspace{1ex}\\
	{{{\bf{K}}_{RL}}}&{{{\bf{K}}_{RR}}}
	\end{array}} \right]\left[ {\begin{array}{*{20}{c}}
	{{\bf{q}}_L^{\left( A \right)}}\vspace{1ex}\\
	{{\bf{q}}_R^{\left( C \right)}}
	\end{array}} \right],
\end{equation}
where ${\bf{K}}$ is the $4\times4$ dynamic stiffness matrix of the deformed unit cell, with its four partitioned $2\times2$ sub-matrices ${{{\bf{K}}_{ij}}} (i,j=L,R)$ being given by
\begin{equation} \label{43}
\begin{array}{l}
{{\bf{K}}_{LL}} = {{\bf{J}}_{LL}} - {{\bf{J}}_{LR}}{\left( {{{\bf{J}}_{RR}} + {\bf{G}}_{LL}^{\left( C \right)}} \right)^{ - 1}}{{\bf{J}}_{RL}},\\
{{\bf{K}}_{LR}} =  - {{\bf{J}}_{LR}}{\left( {{{\bf{J}}_{RR}} + {\bf{G}}_{LL}^{\left( C \right)}} \right)^{ - 1}}{\bf{G}}_{LR}^{\left( C \right)},\\
{{\bf{K}}_{RL}} =  - {\bf{G}}_{RL}^{\left( C \right)}{\left( {{{\bf{J}}_{RR}} + {\bf{G}}_{LL}^{\left( C \right)}} \right)^{ - 1}}{{\bf{J}}_{RL}},\\
{{\bf{K}}_{RR}} =  - {\bf{G}}_{RL}^{\left( C \right)}{\left( {{{\bf{J}}_{RR}} + {\bf{G}}_{LL}^{\left( C \right)}} \right)^{ - 1}}{\bf{G}}_{LR}^{\left( C \right)} + {\bf{G}}_{RR}^{\left( C \right)},
\end{array}
\end{equation}
where
\begin{equation} \label{44}
\begin{array}{l}
{{\bf{J}}_{LL}} = {\bf{G}}_{LL}^{\left( A \right)} - {\bf{G}}_{LR}^{\left( A \right)}{\left( {{\bf{G}}_{RR}^{\left( A \right)} + {\bf{G}}_{LL}^{\left( B \right)}} \right)^{ - 1}}{\bf{G}}_{RL}^{\left( A \right)},\\
{{\bf{J}}_{LR}} =  - {\bf{G}}_{LR}^{\left( A \right)}{\left( {{\bf{G}}_{RR}^{\left( A \right)} + {\bf{G}}_{LL}^{\left( B \right)}} \right)^{ - 1}}{\bf{G}}_{LR}^{\left( B \right)},\\
{{\bf{J}}_{RL}} =  - {\bf{G}}_{RL}^{\left( B \right)}{\left( {{\bf{G}}_{RR}^{\left( A \right)} + {\bf{G}}_{LL}^{\left( B \right)}} \right)^{ - 1}}{\bf{G}}_{RL}^{\left( A \right)},\\
{{\bf{J}}_{RR}} = {\bf{G}}_{RR}^{\left( B \right)} - {\bf{G}}_{RL}^{\left( B \right)}{\left( {{\bf{G}}_{RR}^{\left( A \right)} + {\bf{G}}_{LL}^{\left( B \right)}} \right)^{ - 1}}{\bf{G}}_{LR}^{\left( B \right)}.
\end{array}
\end{equation}

Next, we employ the Bloch-Floquet theorem to derive the dispersion relation of the periodic soft PC plate. Based on the structural periodicity and the incremental continuity conditions at the interface between the neighboring unit cells, the relations of physical quantities at the two ends of each unit cell satisfy
\begin{equation} \label{45}
{\bf{q}}_R^{\left( C \right)} = {{\rm{e}}^{{\rm{i}}\bar k}}{\bf{q}}_L^{\left( A \right)},\quad {\bf{f}}_R^{\left( C \right)} =  - {{\rm{e}}^{{\rm{i}}\bar k}}{\bf{f}}_L^{\left( A \right)},
\end{equation}
where $\bar k=kl$ is the dimensionless Bloch wave number, ranging from $-\pi$ to $\pi$ in the first Brillouin zone, with $k$ being the Bloch wave number. 
Eq.~\eqref{42} combined with Eq.~\eqref{45} leads to
\begin{equation} \label{46}
{{\rm{e}}^{2{\rm{i}}\bar k}}{\bf{q}}_L^{\left( A \right)} + {{\rm{e}}^{{\rm{i}}\bar k}}{\bf{K}}_{LR}^{ - 1}\left( {{{\bf{K}}_{LL}} + {{\bf{K}}_{RR}}} \right){\bf{q}}_L^{\left( A \right)} + {\bf{K}}_{LR}^{ - 1}{{\bf{K}}_{RL}}{\bf{q}}_L^{\left( A \right)} = {{\bf{0}}_{2\times2}}.
\end{equation}
Through some mathematical manipulations, Eq.~\eqref{46} can be rewritten as
\begin{equation} \label{47}
\left[ {\begin{array}{*{20}{c}}
	{{\bf{0}}_{2\times2}}&{{{\bf{I}}_{{\rm{2}} \times {\rm{2}}}}}\\
	{ - {\bf{K}}_{LR}^{ - 1}{{\bf{K}}_{RL}}}&{ - {\bf{K}}_{LR}^{ - 1}\left( {{{\bf{K}}_{LL}} + {{\bf{K}}_{RR}}} \right)}
	\end{array}} \right]\left[ {\begin{array}{*{20}{c}}
	{{\bf{q}}_L^{\left( A \right)}}\\
	{{{\rm{e}}^{{\rm{i}}\bar k}}{\bf{q}}_L^{\left( A \right)}}
	\end{array}} \right] - {{\rm{e}}^{{\rm{i}}\bar k}}\left[ {\begin{array}{*{20}{c}}
	{{\bf{q}}_L^{\left( A \right)}}\\
	{{{\rm{e}}^{{\rm{i}}\bar k}}{\bf{q}}_L^{\left( A \right)}}
	\end{array}} \right] = {{\bf{0}}_{4\times4}},
\end{equation}
where ${\bf{I}}$ is a second-order identity matrix. 

Therefore, by solving the generalized eigenvalue equation \eqref{47}, the dispersion relation of the infinite soft PC plate can be obtained. In the case where a frequency $\omega$ is prescribed, four complex roots of ${{\rm{e}}^{{\rm{i}}\bar k}}$ follow from Eq.~\eqref{47}. 
Say the complex root ${{\rm{e}}^{{\rm{i}}\bar k}}$ is written as $p + {\rm{i}}q$ and the dimensionless Bloch wave number $\bar k$ as $c + {\rm{i}}d$, where $c$, $d$, $p$ and $q$ are all real numbers. Then, we can express $c$ and $d$ in terms of $p$ and $q$ as \citep{han2012modified,zhou2019surface}
\begin{equation} \label{48}
c = \left\{ {\begin{array}{*{20}{l}}
	{\arctan \left( {q/p} \right),}&{{\rm{if}} \quad  p > 0}\\
	{\pi  + \arctan \left( {q/p} \right),}&{{\rm{if}} \quad p < 0 \; \& \; q > 0}\\
	{ - \pi  + \arctan \left( {q/p} \right),}&{{\rm{if}} \quad p < 0 \; \& \; q < 0}
	\end{array}} \right.
\end{equation}
and 
\begin{equation} \label{49}
d = - \frac{{ \ln \left( {{p^2} + {q^2}} \right)}}{2}.
\end{equation}
As a result, the dispersion relation between the frequency and the Bloch wave number for incremental bending waves can be solved from Eqs.~\eqref{47}-\eqref{49}.

%


\subsection{Transmission coefficient of a finite soft dielectric PC plate} \label{Sec4-3}

\begin{figure}[htbp]
	\centering	
	\includegraphics[width=0.85\textwidth]{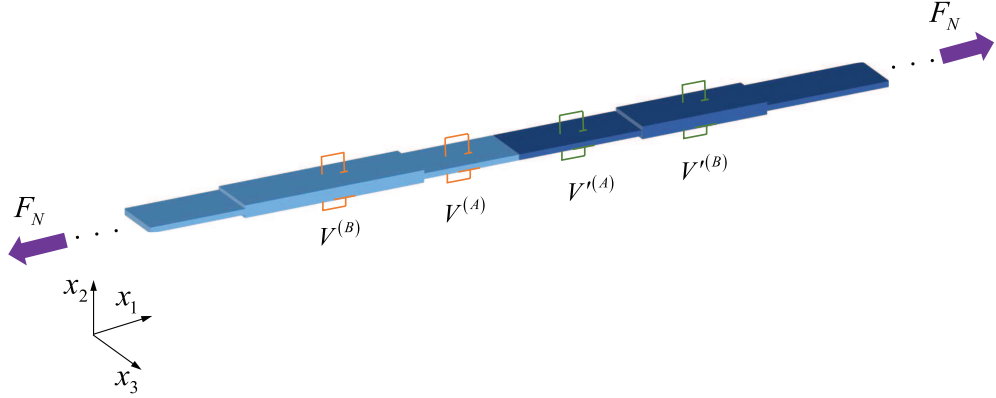}
	\caption{Schematic of a finite soft dielectric PC plate composed of two types of different unit cells under the combination of axial force and electric voltage. The electric voltages applied on the two PC components could be different.}
	\label{Figmixed}
\end{figure}

In this subsection, we analyze the transmission behavior of incremental bending waves propagating in a finite soft dielectric PC plate composed of $N$ unit cells that can be either identical or different (see Fig.~\ref{Figmixed}). In light of Eq.~\eqref{42}, the relation between nodal displacement and force vectors in the $n$-th unit cell can be rewritten as 
\begin{equation} \label{50}
\left[ {\begin{array}{*{20}{c}}
	{{\bf{f}}_{nL}^{\left( A \right)}} \vspace{1ex}\\
	{{\bf{f}}_{nR}^{\left( C \right)}}
	\end{array}} \right] = \left[ {\begin{array}{*{20}{c}}
	{{\bf{K}}_{nLL}^{}}&{{\bf{K}}_{nLR}^{}} \vspace{1ex}\\
	{{\bf{K}}_{nRL}^{}}&{{\bf{K}}_{nRR}^{}}
	\end{array}} \right]\left[ {\begin{array}{*{20}{c}}
	{{\bf{q}}_{nL}^{\left( A \right)}} \vspace{1ex}\\
	{{\bf{q}}_{nR}^{\left( C \right)}}
	\end{array}} \right].
\end{equation}
Using the incremental interfacial continuity conditions between adjacent unit cells, the global stiffness matrix for a finite PC plate with $N$ unit cells can be obtained by assembling the element dynamic stiffness matrix from unit cell 1 to $N$, which yields 
\begin{equation} \label{51}
{{\bf{f}}_{tot}} = {{\bf{Q}}_{tot}}{{\bf{q}}_{tot}},
\end{equation}
where ${{\bf{q}}_{tot}}$, ${{\bf{f}}_{tot}}$ and ${{\bf{Q}}_{tot}}$ are the global displacement vector, global force vector and global stiffness matrix, respectively. The boundary conditions at the two ends of the finite PC plate are also required for calculating the transmission spectrum. If the incident wave is excited at the left end and the output signal is received at the right end of the finite PC plate, we can express the global displacement and force vectors as
\begin{equation} \label{52}
\begin{array}{l}
{{\bf{q}}_{tot}} = \left\{ {\begin{array}{*{20}{c}}
	{{{\bf{q}}^{\left( 0 \right)}}}\\
	{{{\bf{q}}^{\left( 1 \right)}}}\\
	\vdots \\
	{{{\bf{q}}^{\left( N \right)}}}
	\end{array}} \right\}, \quad
{{\bf{f}}_{tot}} = \left\{ {\begin{array}{*{20}{c}}
	{{{\bf{f}}^{\left( 0 \right)}}}\\
	{\bf{0}}\\
	\vdots \\
	{\bf{0}}
	\end{array}} \right\}, \vspace{1ex}\\
{{\bf{f}}^{\left( 0 \right)}} = \left\{ {\begin{array}{*{20}{c}}
	{{0}}\\
	F_I
	\end{array}} \right\}, \quad
{{\bf{q}}^{\left( 0 \right)}} = \left\{ {\begin{array}{*{20}{c}}
	{{u_{2I}}}\\
	{{\varphi _I}}
	\end{array}} \right\}, \quad
{{\bf{q}}^{\left( N \right)}} = \left\{ {\begin{array}{*{20}{c}}
	{{u_{2O}}}\\
	{{\varphi _O}}
	\end{array}} \right\},
\end{array}
\end{equation}
where  ${\bf{q}}^{\left( n \right)}={{\bf{q}}_{nR}^{\left( C \right)}}={{\bf{q}}_{(n+1)L}^{\left( A \right)}}$ and ${\bf{f}}^{\left( n \right)}={{\bf{f}}_{nR}^{\left( C \right)}}+{{\bf{f}}_{(n+1)L}^{\left( A \right)}}$ are the nodal displacement and external force vectors between adjacent unit cells, $F_I$ is the shear force excited at the input end, and $u_{2I}$ ($\varphi _I$) and $u_{2O}$ ($\varphi _O$) are the displacement (rotation angle) signals at the input and output ends, respectively, which are caused by the input shear force. 
In particular, we have ${\bf{q}}^{\left( 0 \right)}={{\bf{q}}_{1L}^{\left( A \right)}}$,  ${\bf{f}}^{\left( 0 \right)}={{\bf{f}}_{1L}^{\left( A \right)}}$, ${\bf{q}}^{\left( N \right)}={{\bf{q}}_{NR}^{\left( C \right)}}$ and ${\bf{f}}^{\left( N \right)}={{\bf{f}}_{NR}^{\left( C \right)}}$. 
The elements ${\bf{f}}^{\left( n \right)}$ of the global force vector ${{\bf{f}}_{tot}}$ in Eq.~\eqref{52} are all equal to zero, except for ${\bf{f}}^{\left( 0 \right)}$ at the incident end because of the traction-free surfaces and input excitation of shear force. After assembly, the global stiffness matrix is written as
\begin{small}
\begin{equation} \label{53}
\begin{array}{l}
{{\bf{Q}}_{tot}} =\\[4pt]
 \left[ {\begin{array}{*{20}{c}}
	{{\bf{K}}_{LL}^{\left( 1 \right)}}&{{\bf{K}}_{LR}^{\left( 1 \right)}}&{\bf{0}}&{\bf{0}}& \cdots &{\bf{0}}&{\bf{0}}\\
	{{\bf{K}}_{RL}^{\left( 1 \right)}}&{{\bf{K}}_{RR}^{\left( 1 \right)} + {\bf{K}}_{LL}^{\left( 2 \right)}}&{{\bf{K}}_{LR}^{\left( 2 \right)}}&{\bf{0}}& \vdots & \vdots &{\bf{0}}\\
	{\bf{0}}&{{\bf{K}}_{RL}^{\left( 2 \right)}}&{{\bf{K}}_{RR}^{\left( 2 \right)} + {\bf{K}}_{LL}^{\left( 3 \right)}}& \cdots & \cdots & \cdots & \vdots \\
	{\bf{0}}&{\bf{0}}& \vdots & \ddots & \vdots &{\bf{0}}&{\bf{0}}\\
	\vdots & \cdots & \cdots & \cdots &{{\bf{K}}_{RR}^{\left( {N - 2} \right)} + {\bf{K}}_{LL}^{\left( {N - 1} \right)}}&{{\bf{K}}_{LR}^{\left( {N - 1} \right)}}&{\bf{0}}\\
	{\bf{0}}& \cdots & \ddots &{\bf{0}}&{{\bf{K}}_{RL}^{\left( {N - 1} \right)}}&{{\bf{K}}_{RR}^{\left( {N - 1} \right)} + {\bf{K}}_{LL}^{\left( N \right)}}&{{\bf{K}}_{LR}^{\left( N \right)}}\\
	{\bf{0}}&{\bf{0}}&{\bf{0}}&{\bf{0}}&{\bf{0}}&{{\bf{K}}_{RL}^{\left( N \right)}}&{{\bf{K}}_{RR}^{\left( N \right)}}
	\end{array}} \right].
\end{array}
\end{equation}
\end{small}

Because the global force vector ${{\bf{f}}_{tot}}$ is known when $F_I$ is fixed, the displacements $u_{2I}$ and $u_{2O}$ at input and output ends can be solved according to Eq.~\eqref{51}. The transmission coefficient can then be defined and calculated as
\begin{equation} \label{54}
t_N = 20\log \left( {\frac{{\left| {{u_{2O}}} \right|}}{{\left| {{u_{2I}}} \right|}}} \right).
\end{equation}
%

\section{Numerical results and discussion}\label{section5}


In this section, we present numerical results to analyze the tunable effects of the applied axial force and electric voltage on the dispersion relation and topological properties of incremental bending waves for the soft PC plate described by the Gent ideal dielectric model. 

In the following numerical simulations, we set the geometric parameters of the undeformed unit cell as: 
${L^{\left( A \right)}} ={{L\left( {1 + \delta } \right)}}/2$ and 
${H^{\left( A \right)}} = 1{\text{ cm}}$ for the length and thickness for sub-plate $A$; for sub-plate $B$, the length is ${L^{\left( B \right)}} ={{L\left( {1 - \delta } \right)}}/2$ with the thickness being ${H^{\left( B \right)}} = 3{\text{ cm}}$, where ${L} = 15{\text{ cm}}$ is the total length of unit cell and $\delta$ is a structural parameter ranging from $-1$ to 1. 
For the commercial product Fluorosilicone 730 \citep{pelrine2000high}, the initial density, shear modulus and relative permittivity of soft dielectric PC plate are $\rho  = 1400{\text{ kg}}/{{\text{ m}}^3}$,  
$\mu  = 167.67$ kPa and
${\varepsilon _r} = 7.11$, respectively. We define the dimensionless axial force and voltage as 
${{\overline{F}}_{N}}={F_N}/(\mu w{H^{\left( B \right)}})$ and
${{\overline V}^{\left( p \right)}} = {V^{\left( p \right)}}\sqrt {\varepsilon /\mu } /{H^{\left( B \right)}}$, respectively. The ordinary frequency $f$, which is measured in Hz, is related to the circular frequency by $f=\omega/(2\pi)$. Moreover, the width $w$ can be eliminated in the process of parameter nondimensionalization and thus its specific value is not needed here.


\subsection{Nonlinear static response of soft dielectric PC plate} \label{Sec5-1}

When subjected to the axial force and voltage, the finite static deformation of the incompressible dielectric PC plate is considered first. 

\begin{figure}[htbp]
	\centering
	\setlength{\abovecaptionskip}{5pt}
	\setlength{\belowcaptionskip}{0pt}
	\includegraphics[width=0.32\textwidth]{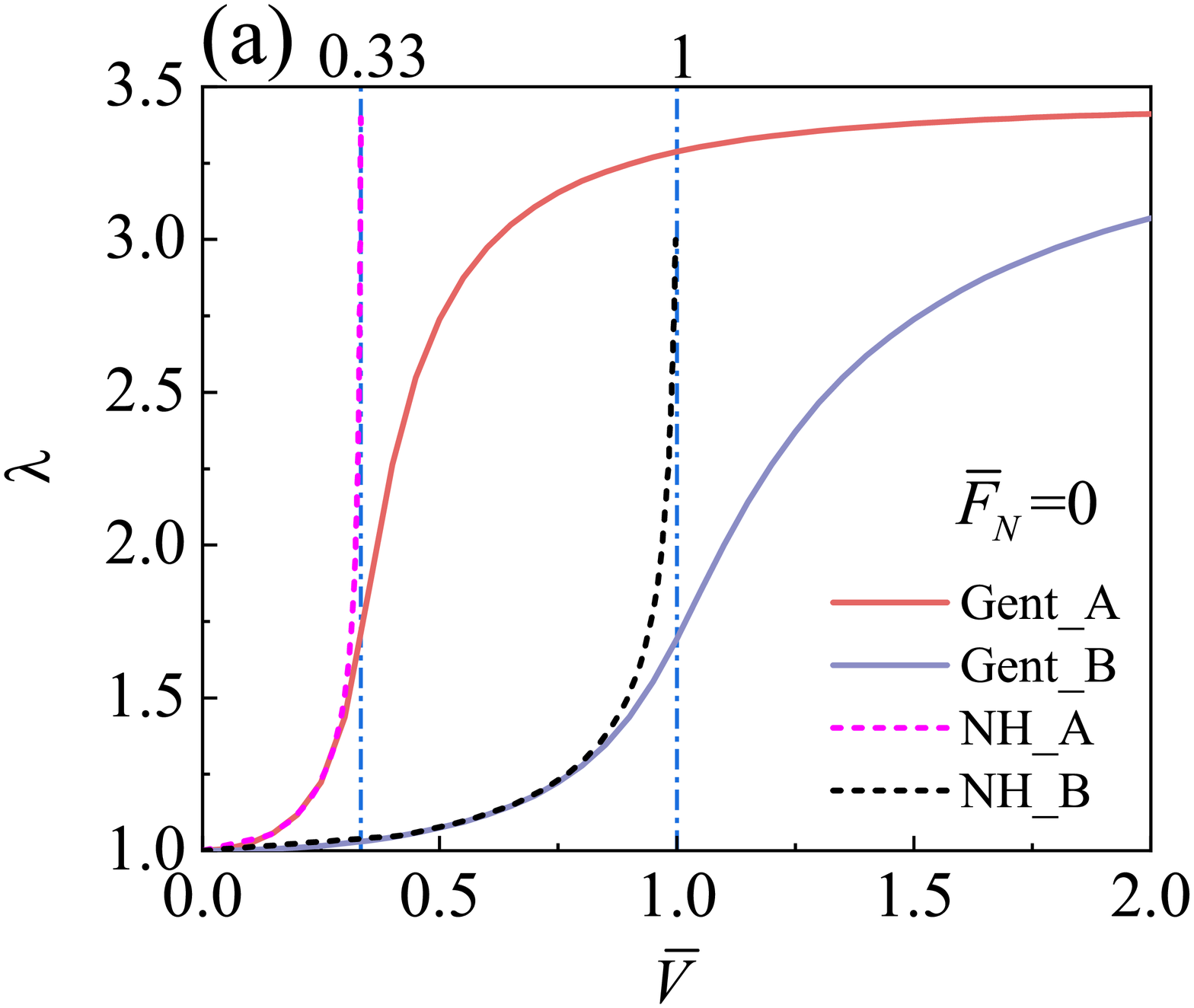}
	\includegraphics[width=0.32\textwidth]{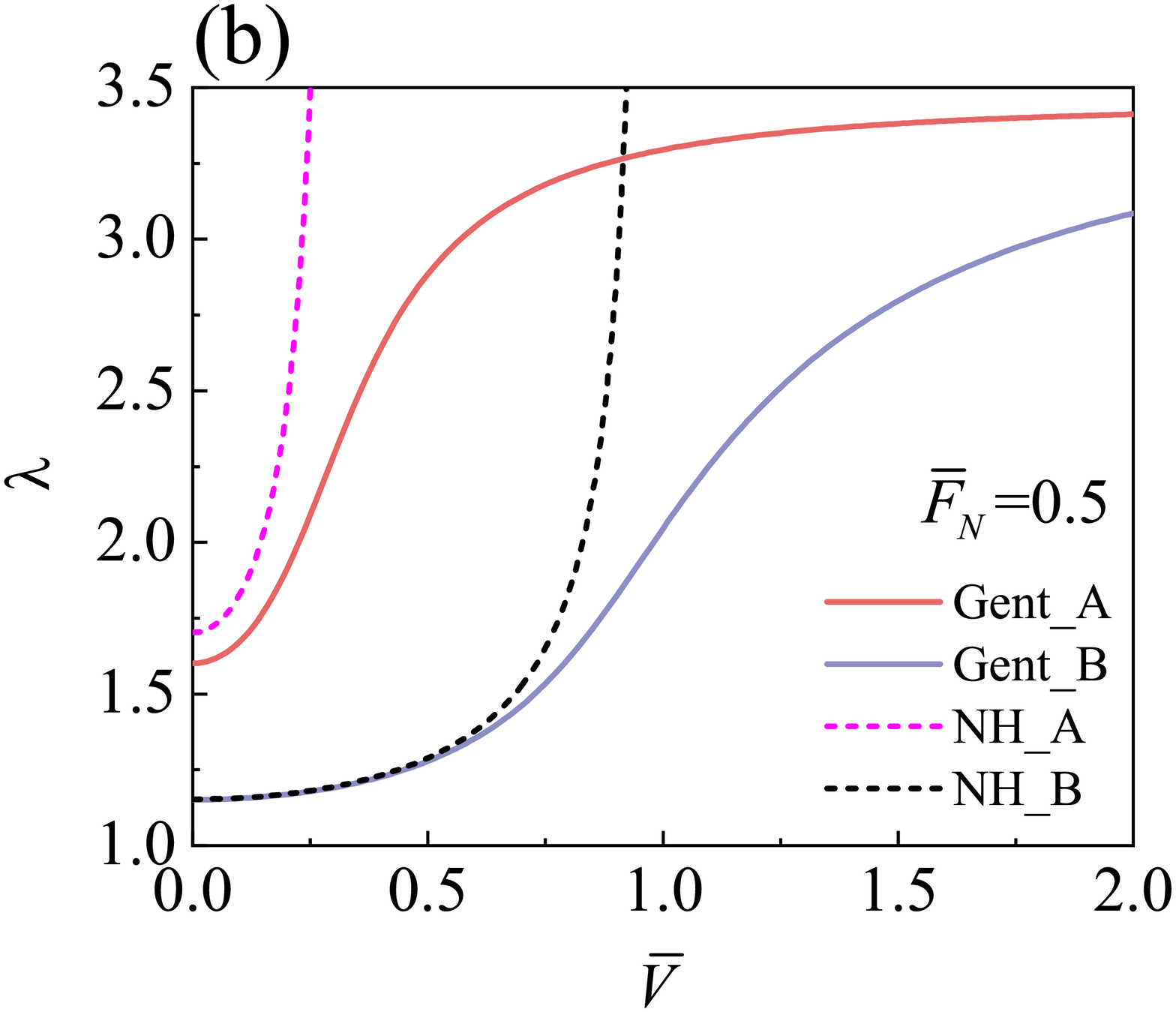}
	\includegraphics[width=0.32\textwidth]{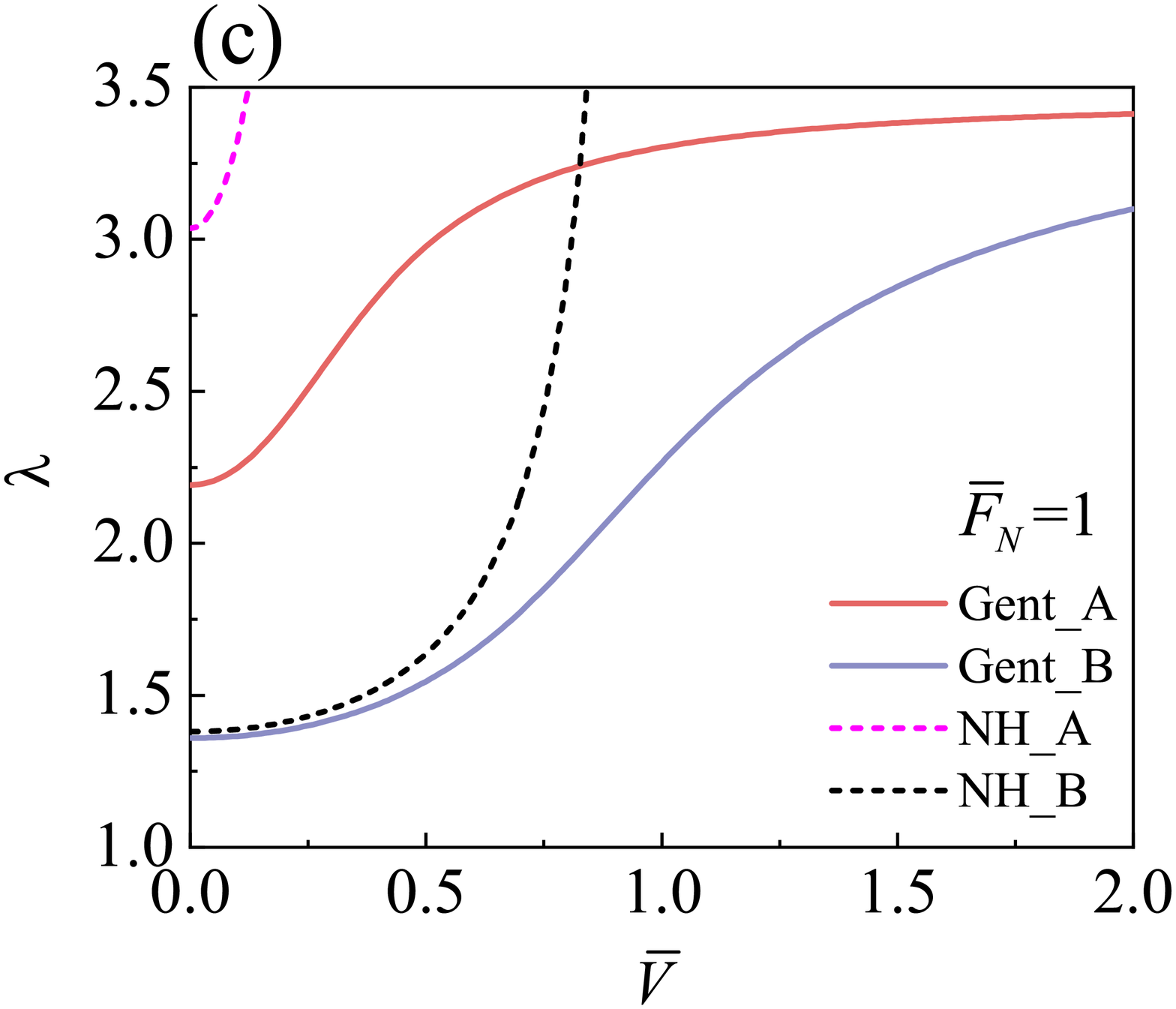}
	\caption{Nonlinear response of stretch ratios $\lambda^{\left( A \right)}$ and $\lambda^{\left( B \right)}$ to the normalized voltage ${{\overline{V}}}$ in the soft dielectric PC plate subjected to various axial forces: (a) ${{\overline{F}}_{N}}=0$; (b) ${{\overline{F}}_{N}}=0.5$; (c) ${{\overline{F}}_{N}}=1$ for the Gent model ($J_{m}=10$) and neo-Hookean model.}
	\label{Fig2}
\end{figure}

In the absence of the axial force ${{\overline{F}}_{N}}=0$, Fig.~\ref{Fig2}(a) presents the stretch ratios $\lambda^{\left( A \right)}$ and $\lambda^{\left( B \right)}$ as functions of the dimensionless voltage ${{\overline V}}$ for the ideal dielectric Gent ($J_m=10$) and neo-Hookean ($J_m \to \infty$) models.
For the dielectric neo-Hookean PC plate when ${{\overline{F}}_{N}}=0$, Eq.~\eqref{20}  can be 
simplified as
\begin{equation} \label{55}
	{\overline V^{\left( p \right)}} = \sqrt {\left[ {1 - {{\left( {{\lambda ^{\left( p \right)}}} \right)}^{ - 4}}} \right]} {\overline H^{\left( p \right)}},
\end{equation}
where ${\overline H^{\left( p \right)}}={H^{\left( p \right)}}/{H^{\left( B \right)}}$.
Note from Eq.~\eqref{55} that the adjusting range of the applied voltage ${\overline V^{\left( p \right)}}$ is confined and smaller than the limit value ${\overline H^{\left( p \right)}}$, which is in agreement with the result shown in Fig.~\ref{Fig2}(a). 
As the applied voltage increases to $\overline H^{\left( p \right)}$, the axial stretch ratio increases rapidly, and the electromechanical instability or pull-in instability may emerge for the soft dielectric PC plate described by the neo-Hookean model. 
Clearly, the axial deformation of the thinner plate (component $A$) is larger than that of the thicker plate (component $B$) when subjected to the same electric load. 
For the Gent PC plate, the stretch ratio keeps increasing with the voltage and then reaches a plateau at the lock-up stretch.  
Note that when the applied axial force increases (such as when ${{\overline{F}}_{N}}=0.5$ and 1 in Figs.~\ref{Fig2}(b) and (c)), a larger initial stretch at ${{\overline{V}}}=0$ can be obtained, and the deformation of the Gent phase is smaller than that of the neo-Hookean phase at the same level of electric voltage. 
For the neo-Hookean plate, the difference in the stretch ratios of sub-plates $A$ and $B$ becomes much larger than that for the Gent case when increasing the axial force.
Due to the limited range for adjusting the electric voltage in the neo-Hookean PC, we focus on the soft dielectric PC plate described by the Gent model ($J_m=10$) in the following section.

%
 
 
 \subsection{Tunable effect of the electric voltage on bending waves} \label{Sec5-2}
 
 
 Based on the nonlinear deformation generated by the external loads, we now discuss the band structure and tunable topological interface states of the superimposed bending waves in the dielectric PC plate.

\begin{figure}[h!]
 	\centering
 	\setlength{\abovecaptionskip}{5pt}
 	\includegraphics[width=0.3\textwidth]{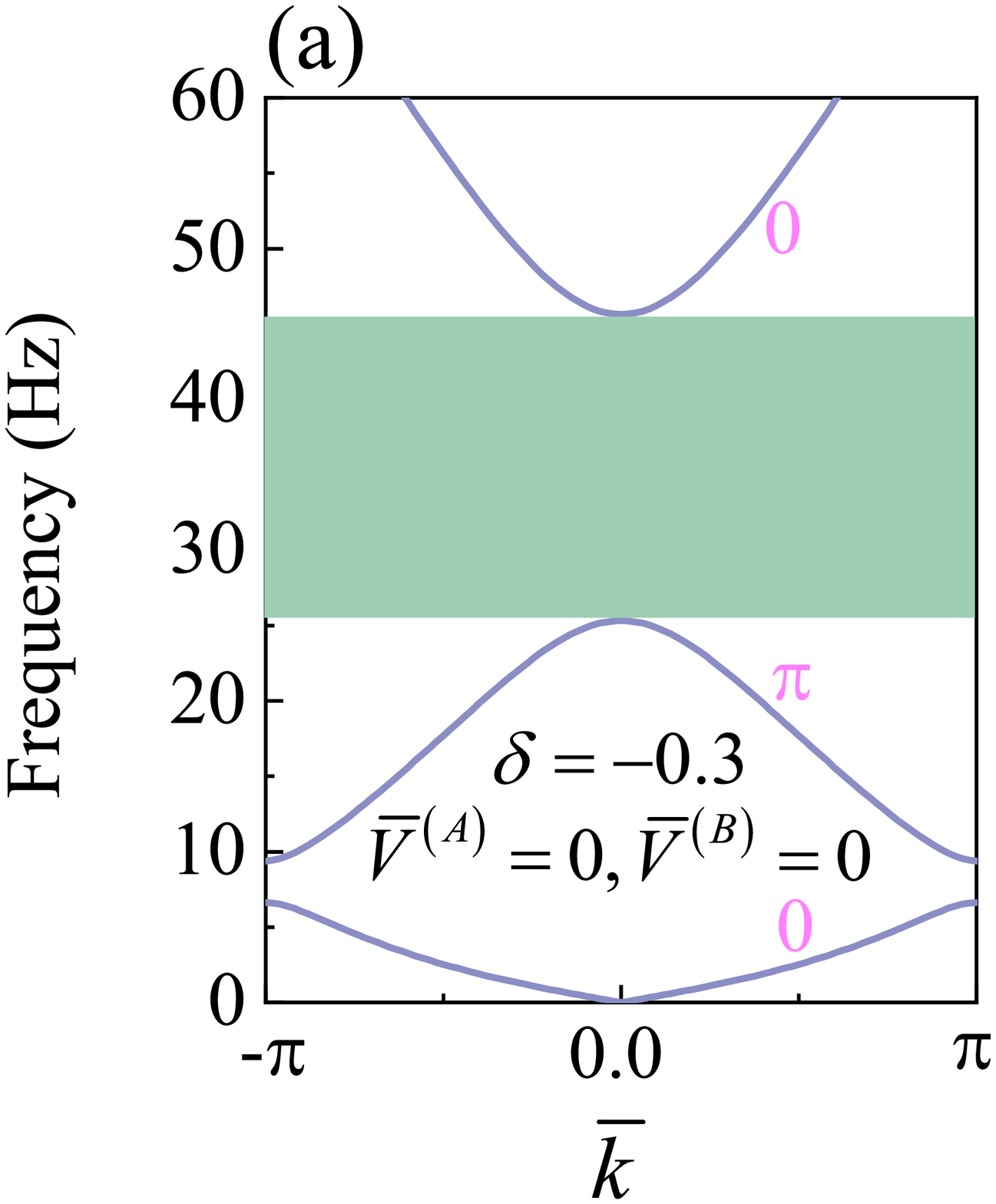}
 	\includegraphics[width=0.3\textwidth]{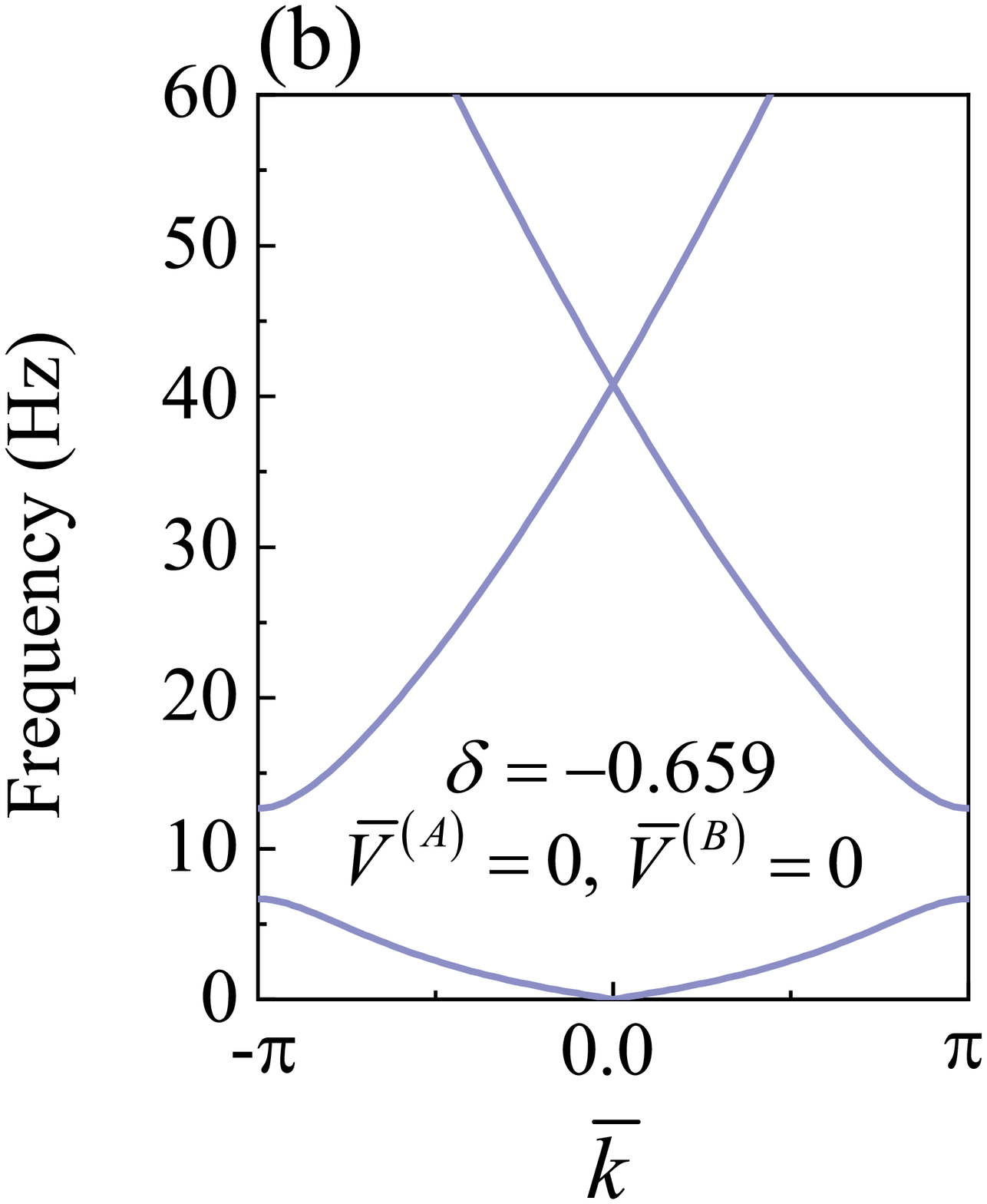}
 	\includegraphics[width=0.3\textwidth]{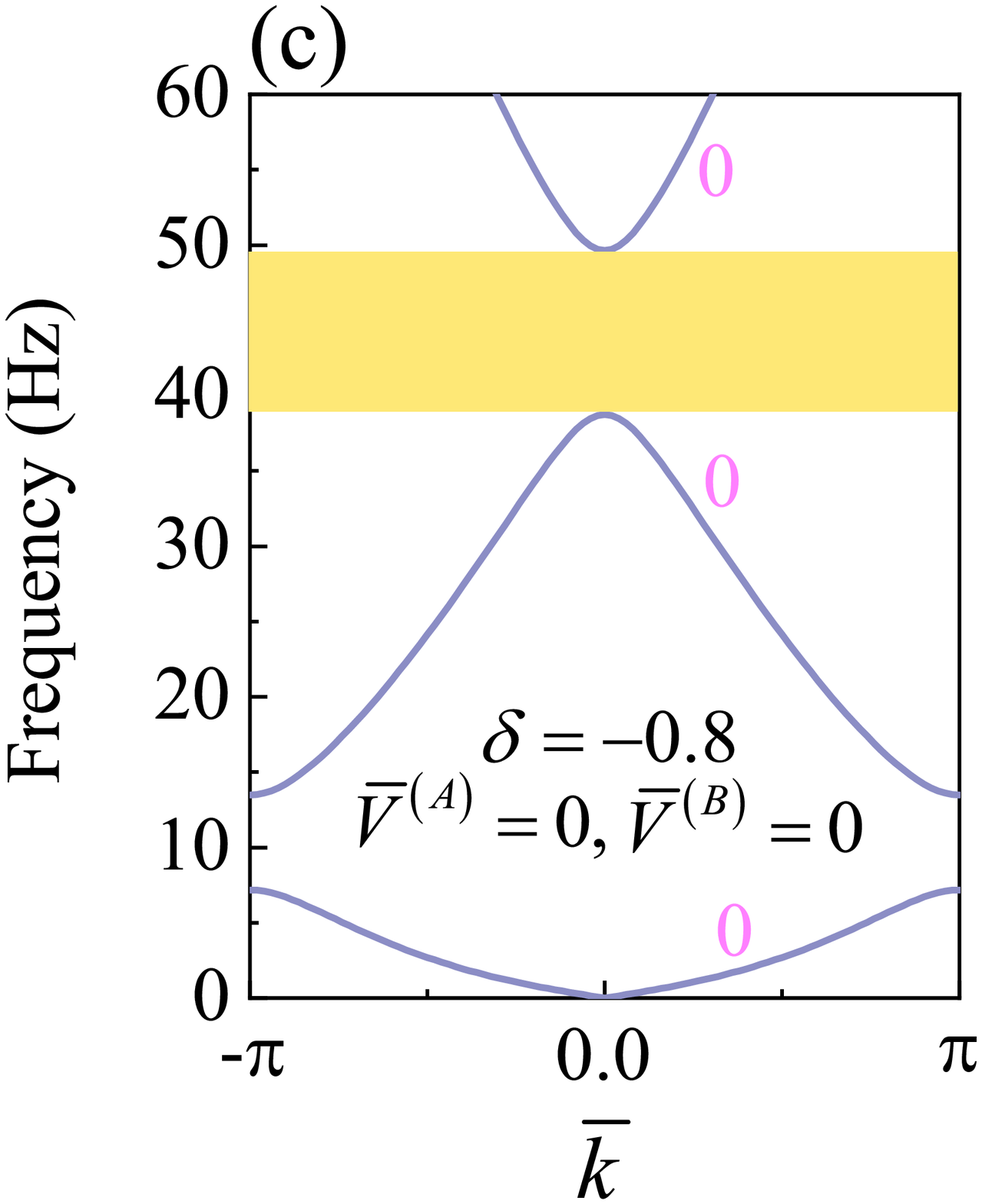}
 	\includegraphics[width=0.3\textwidth]{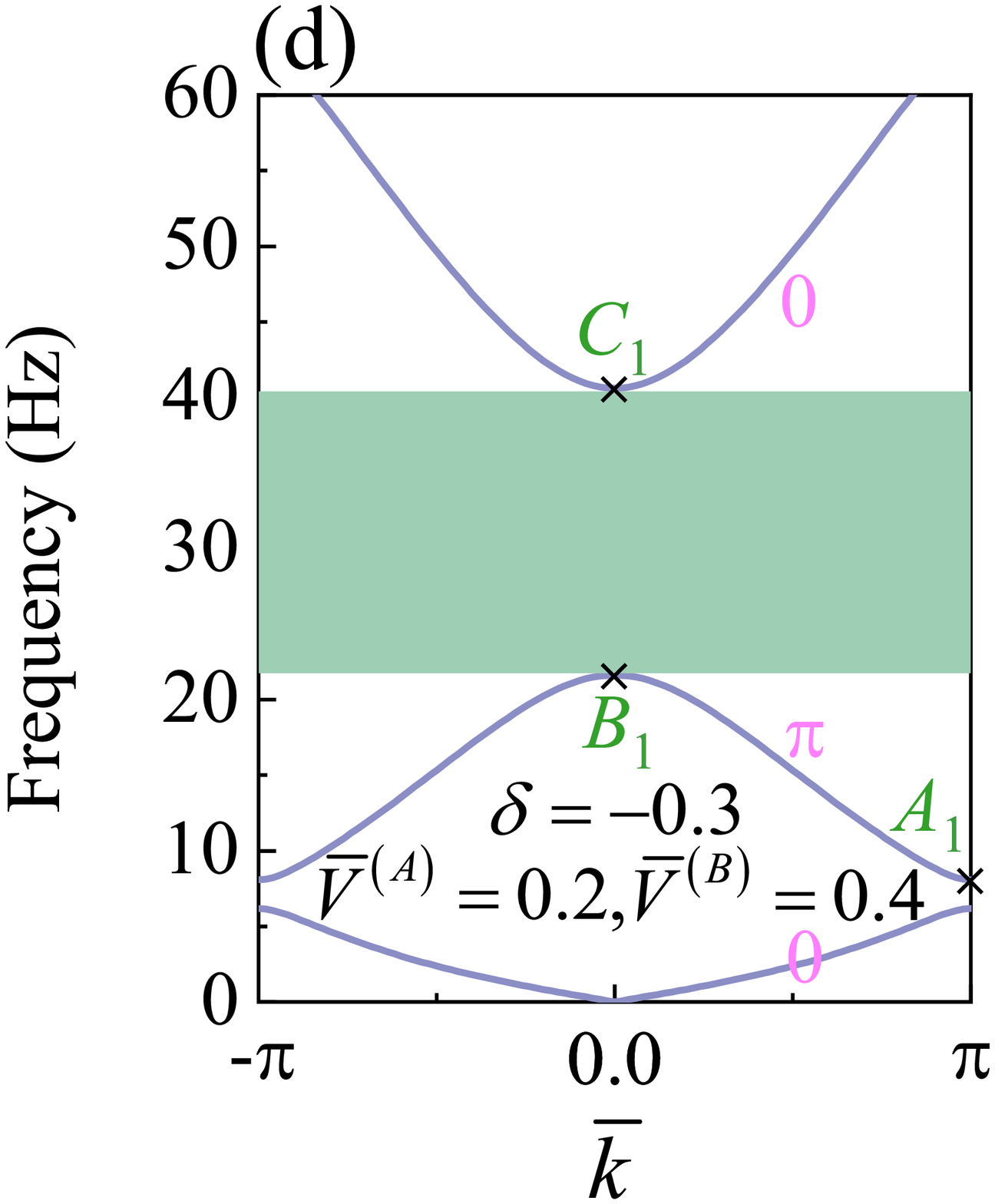}
 	\includegraphics[width=0.3\textwidth]{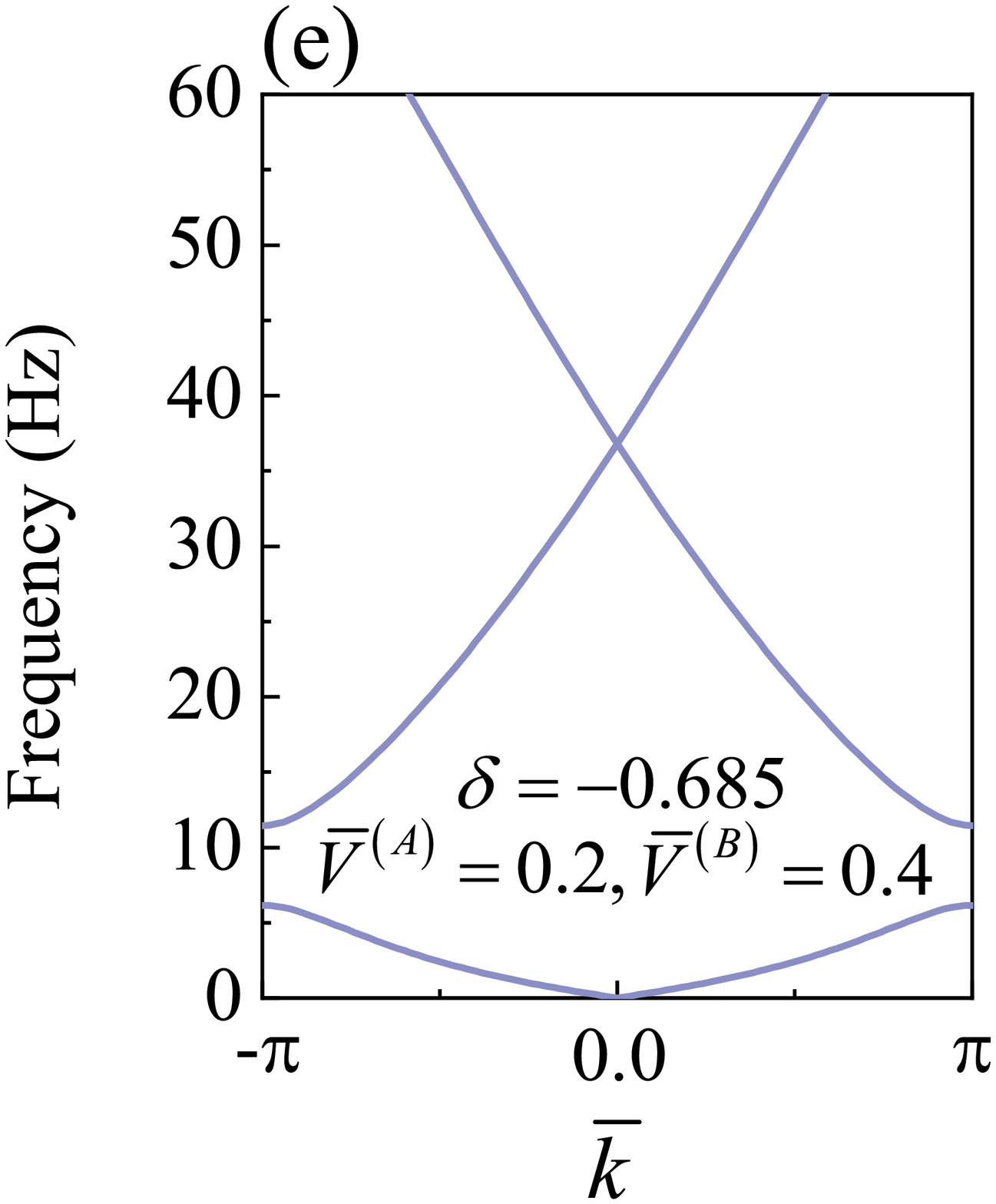}
 	\includegraphics[width=0.3\textwidth]{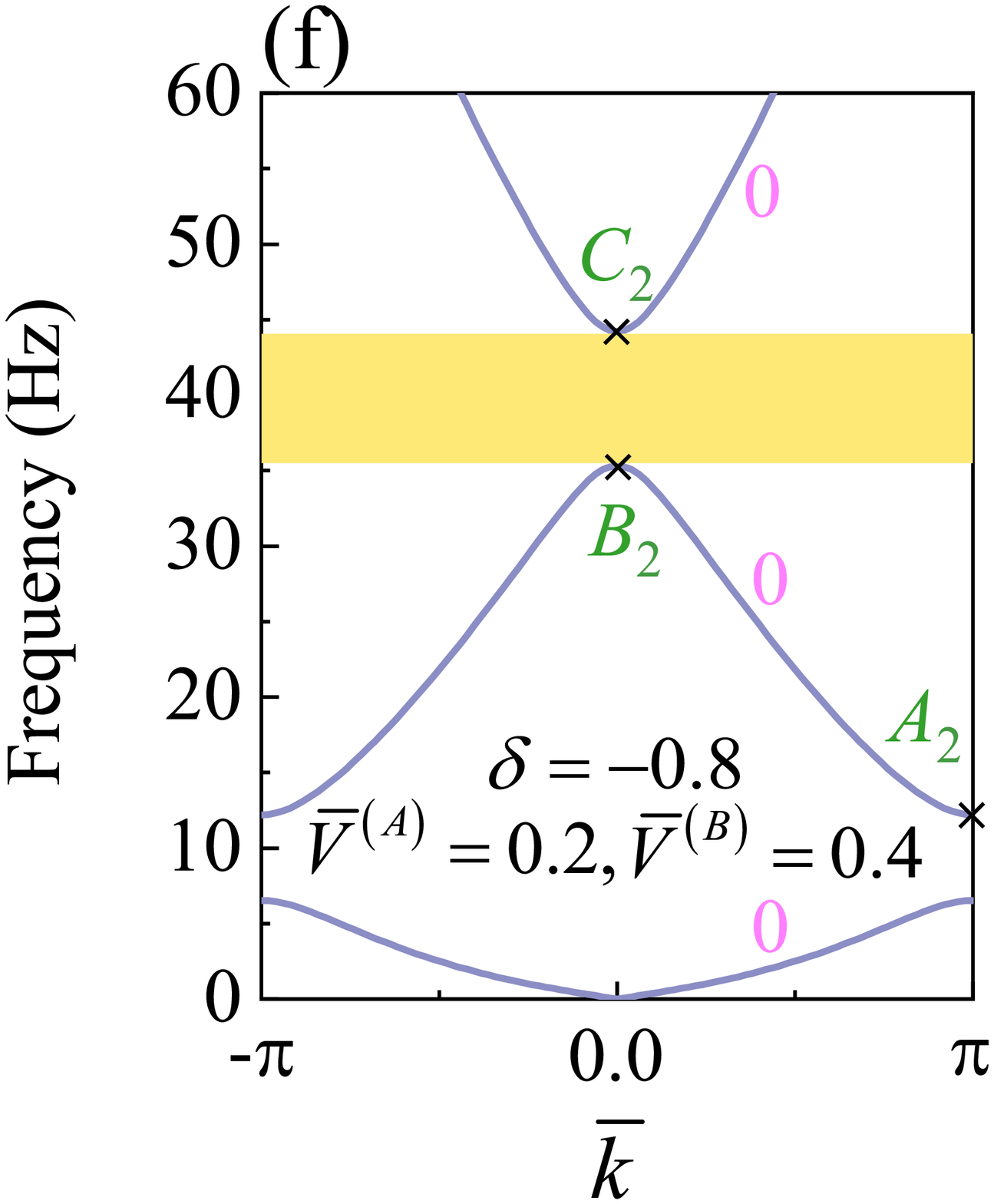}
 	\caption{Band structures of bending waves in the dielectric Gent ($J_m=10$) PC plate without axial force (${{\overline{F}}_{N}}=0$) for different electric voltages and initial geometrical parameter $\delta$: (a)-(c) Topological transition process in the absence of electric voltage (${{\overline{V}}^{(A)}}={\overline{V}}^{(B)}=0$) for three different values of ${\delta}=-0.3$,  $-0.659$ and  $-0.8$, respectively; (d)-(f) Topological transition process for ${{\overline{V}}^{(A)}}=0.2$ and ${\overline{V}}^{(B)}=0.4$ with three different values of ${\delta}=-0.3$,  $-0.685$ and $-0.8$. The Zak phases of the first three bands are marked by magenta font. The stripes in green and yellow indicate the second BG signs with $\varsigma < 0$ and $\varsigma >0$, respectively.}
 	\label{Fig3}
 \end{figure}

 For the bending waves, the band structures of the dielectric Gent PC plate in the absence of axial force (${{\overline{F}}_{N}}=0$) are shown in Fig.~\ref{Fig3} for different initial structural parameters $\delta$ and electric voltages $\overline{V}$. Figs.~\ref{Fig3}(a)-(c) illustrate the band inversion process for ${{\overline{V}}^{(A)}}={\overline{V}}^{(B)}=0$, and the corresponding results for ${{\overline{V}}^{(A)}}=0.2$ and ${\overline{V}}^{(B)}=0.4$ are displayed in Figs.~\ref{Fig3}(d)-(f). 
 Here, the soft dielectric PC plates with different $\delta$ stand for distinct unit-cell configurations with identical initial unit-cell length.

We observe from Figs.~\ref{Fig3}(a)-(c) that with the decrease of $\delta$ from $-0.3$ to $-0.659$, the second BG for ${{\overline{V}}^{(A)}}={\overline{V}}^{(B)}=0$ closes at the center of the Brillouin zone, where a linear crossover, termed as the Dirac cone, occurs and marks a topological transition point. We can see that the occurrence of BG degeneracy at $\delta=-0.659$ corresponds to a case where sub-plates $A$ and $B$ are not equally divided in the unit cell, a notable difference from the corresponding result for longitudinal waves \citep{chen2021low}. The second BG may reopen when decreasing $\delta$ further from $-0.659$ to $-0.8$. Hence, varying the geometrical parameter $\delta$ can result in the evolution process that the second BG opens, closes and reopens. This band inversion process is related to the exchange of topological phase, and will be explained later in details. In Figs.~\ref{Fig3}(d)-(f), a similar topological transition process for ${{\overline{V}}^{(A)}}=0.2$ and ${\overline{V}}^{(B)}=0.4$ can be also observed. Compared with the results corresponding to ${{\overline{V}}^{(A)}}={\overline{V}}^{(B)}=0$, the topological transition point is located at a different $\delta=-0.685$ with its frequency lowered, which is due to the increasing deformation induced by the electric load.

To describe the topological properties, we invoke the Zak phase, a concept which originates from electronic systems science and is a special type of Berry phase for the 1D case \citep{xiao2015geometric, yin2018band}. By summing the Zak phases of all bulk bands below one BG, we arrive at the topological property of this BG. 
It is known that the value of Zak phase also depends on the choice of a unit cell: if the unit cell is arbitrary without mirror symmetry, then the Zak phase can take any value \citep{feng2019magnetically}; if mirror symmetry of the unit cell (see Fig.~\ref{Fig1}(a) and (b)) is ensured, then the Zak phase can only be calculated as 0 or $\pi$. 
For the soft dielectric PC plate subjected to external loads, this conclusion is still valid. 
Here, we denote the two PC configurations with $\delta=-0.3$ and $-0.8$ as the S1 and S2 configurations for simplicity. 

\begin{figure}[htbp]
	\centering
	\setlength{\abovecaptionskip}{5pt}
	\includegraphics[width=0.32\textwidth]{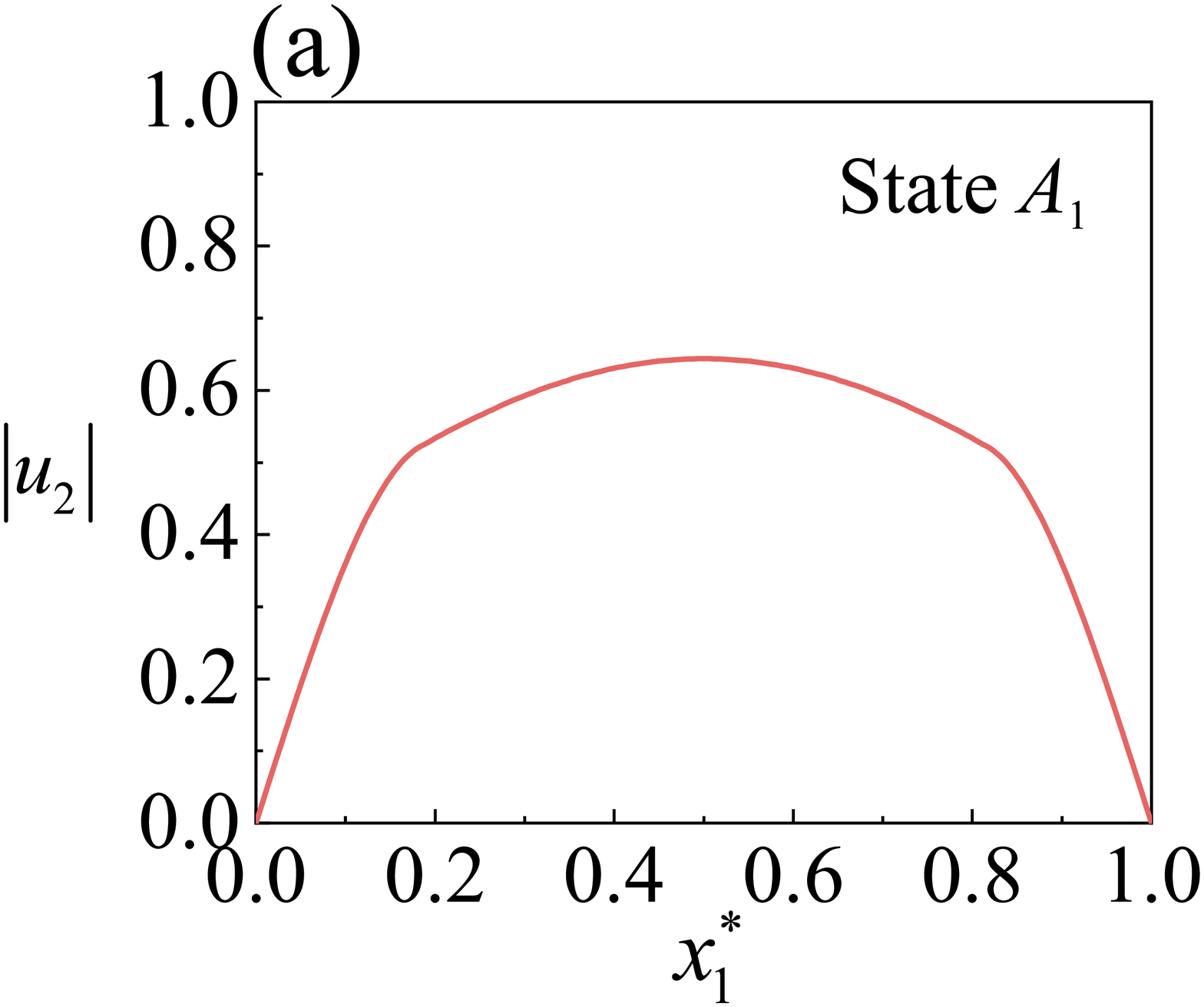}
	\hspace{0.002\textwidth}
	\includegraphics[width=0.32\textwidth]{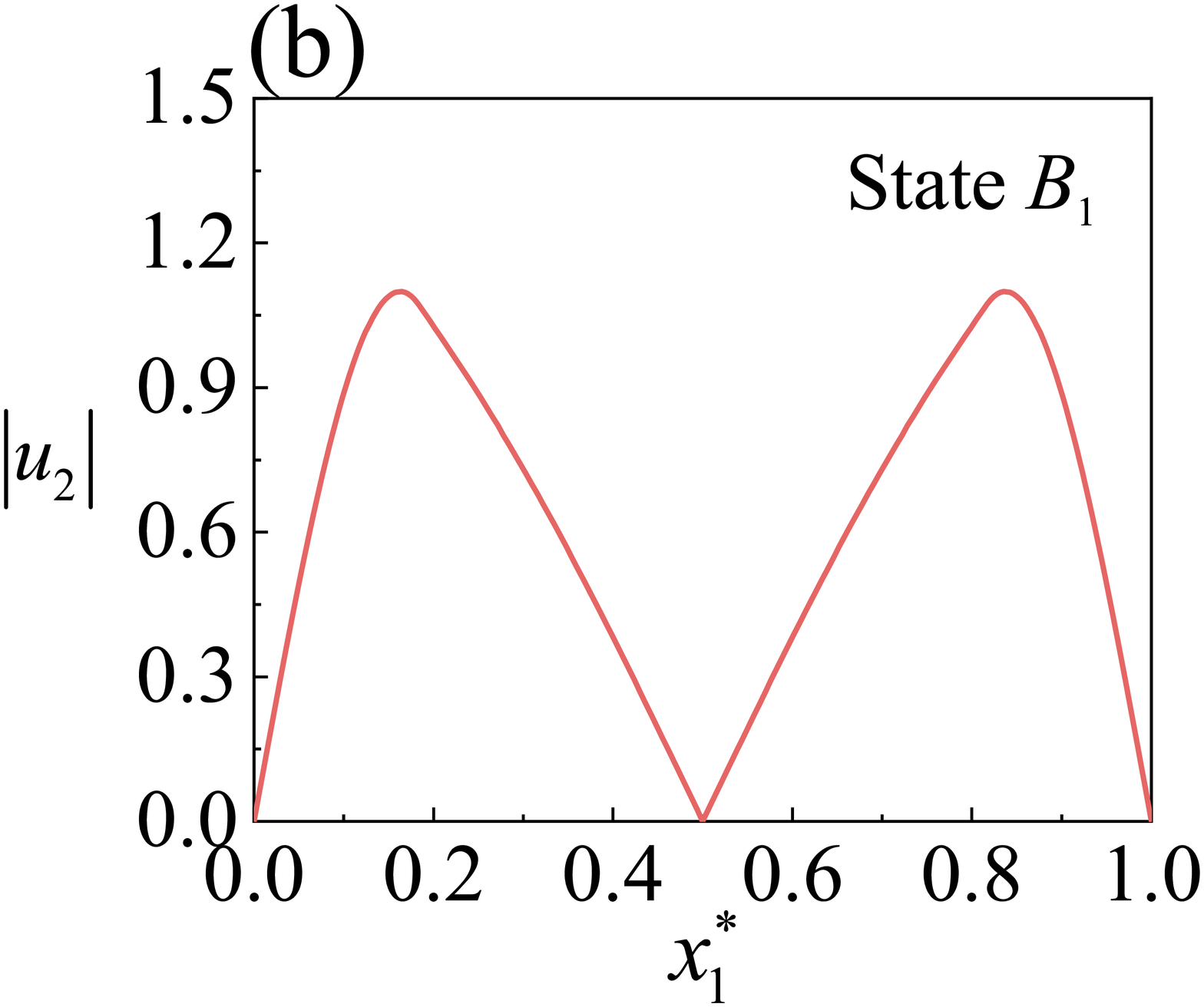}
	\hspace{0.002\textwidth}
	\includegraphics[width=0.32\textwidth]{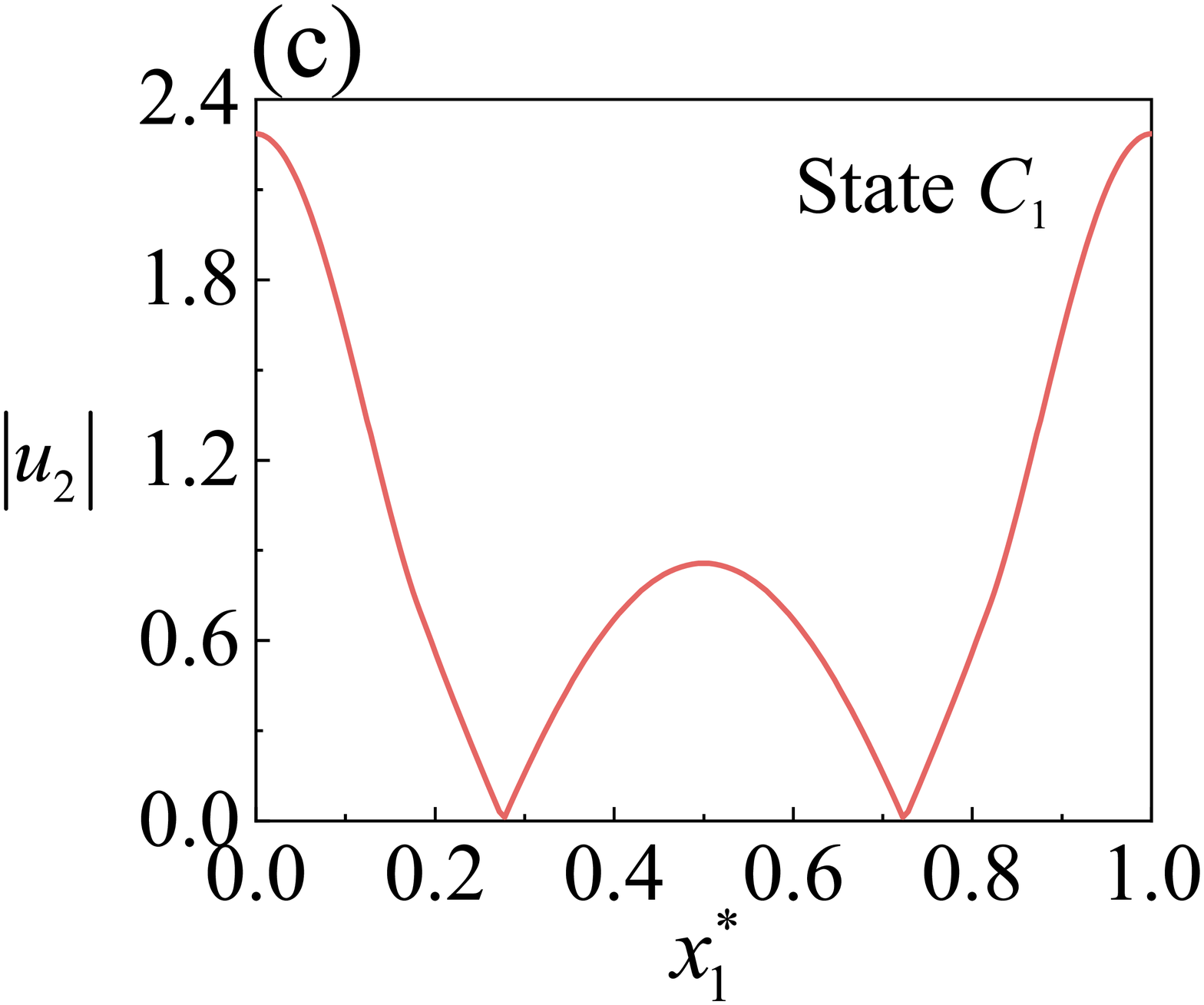}
	\includegraphics[width=0.32\textwidth]{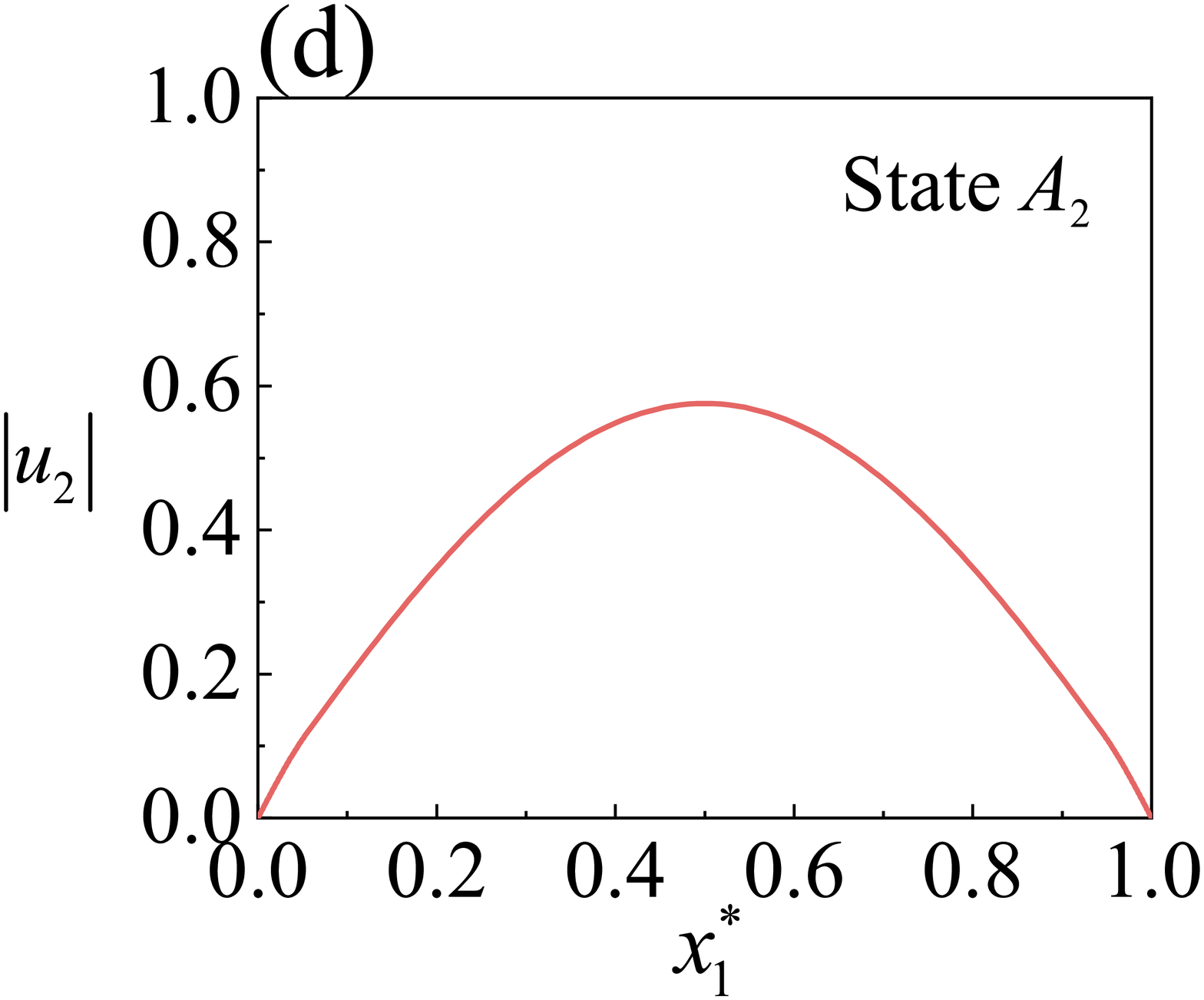}
	\hspace{0.002\textwidth}
	\includegraphics[width=0.32\textwidth]{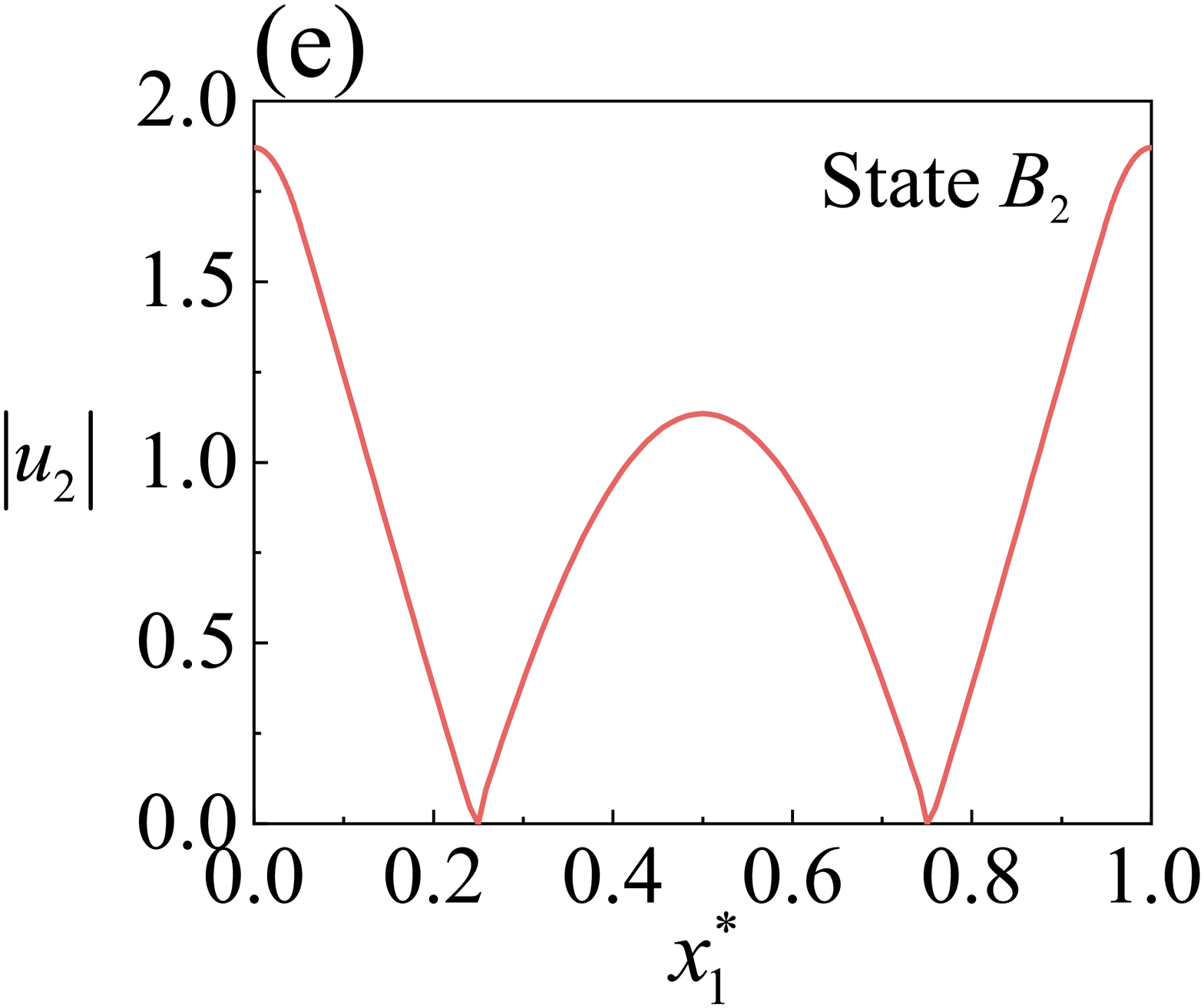}
	\hspace{0.002\textwidth}
	\includegraphics[width=0.32\textwidth]{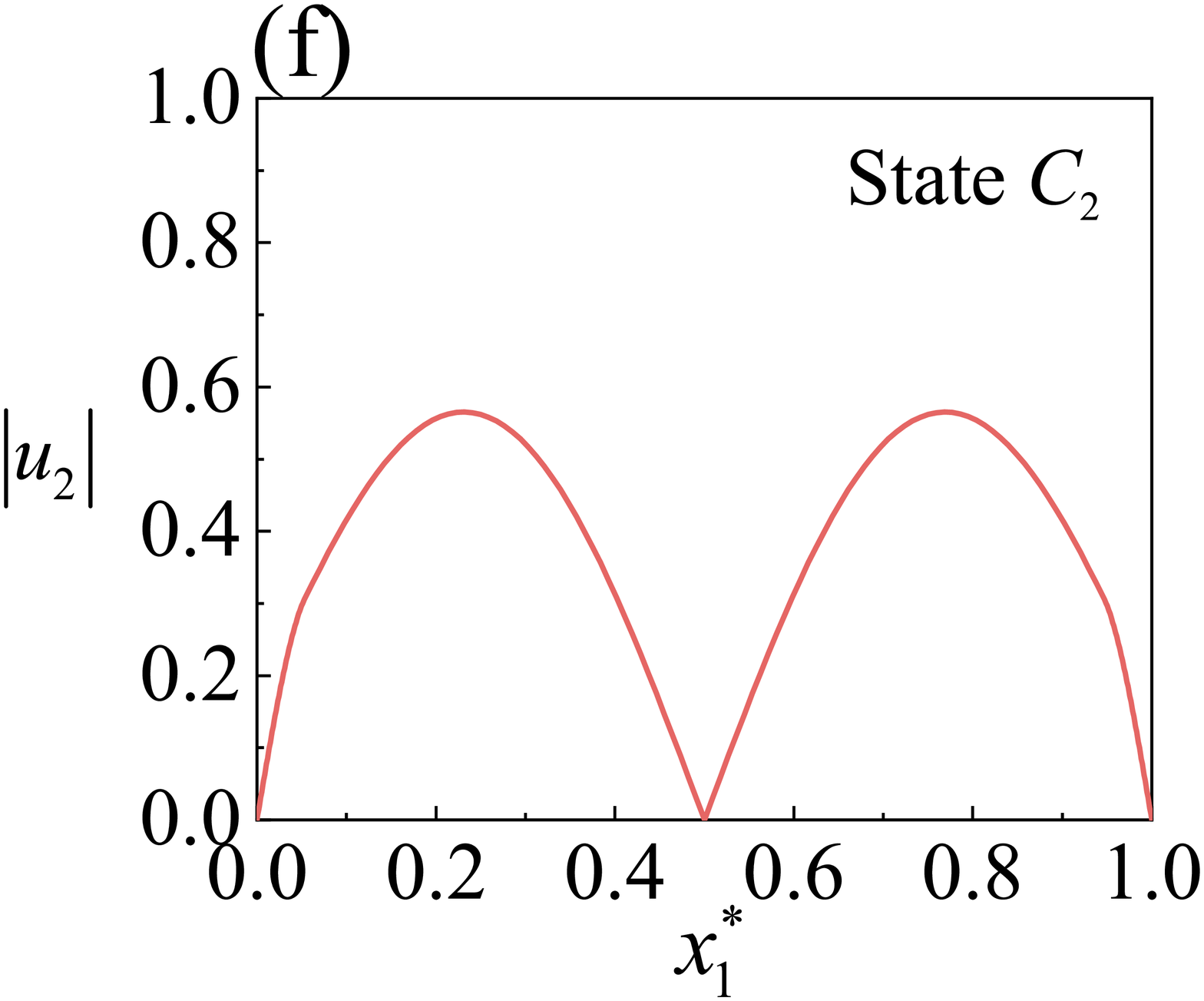}
	\caption{The absolute value of deflection $u_2$ of six band-edge states as a function of the normalized axial position  ${x_1^{*}} = {x_1}/{l}$ in the deformed Gent unit cell ($J_m=10$) for ${{\overline{V}}^{(A)}}=0.2$ and ${\overline{V}}^{(B)}=0.4$. Band-edge states $A_1$, $B_1$, $C_1$, $A_2$, $B_2$, $C_2$ correspond to the cross symbols in Figs.~\ref{Fig3}(d) and (f), and are illustrated in Figs.~\ref{Fig4}(a)-(f), respectively.}
	\label{Fig4}
\end{figure}

To obtain the Zak phase of a bulk band of the 1D PC system, we examine the symmetry properties \citep{kohn1959analytic, xiao2014surface, chen2021low} of edge states at the center and border of the Brillouin zone. The absolute values of deflection $u_2$ of six band-edge states $A_1$, $B_1$, $C_1$, $A_2$, $B_2$, $C_2$, which correspond to the cross marks shown in Figs.~\ref{Fig3}(d) and (f), are calculated. For the deformed Gent unit cell subjected to ${{\overline{V}}^{(A)}}=0.2$ and ${\overline{V}}^{(B)}=0.4$, the spatial distributions of $\left| {{u_2}} \right|$ as a function of the normalized coordinate ${x_1^{*}} = {x_1}/{l}$ are displayed in Fig.~\ref{Fig4} for different band-edge states.  
If the band-edge states at two symmetry points (center and border) of the Brillouin zone for the same bulk band have different symmetries, then the Zak phase of this band is $\pi$. 
Otherwise, the Zak phase is 0. 
 
Based on the symmetric property, the unit-cell displacement field can be divided into even and odd eigenmodes. 
For the even eigenmode (see Fig.~\ref{Fig4}(a)), the displacement amplitude at the center of unit cell is nonzero; for the odd eigenmode (see Fig.~\ref{Fig4}(b)), the amplitude of $u_2$ at the unit-cell center is zero.
As we can see, edge states $A_1$ and $B_1$ possess opposite symmetric properties with respect to the unit cell in Configuration S1, and the related Zak phase of the second band is $\pi$. 
When the geometrical parameter is changed, band-edge states $A_2$ and $B_2$ in the S2 Configuration have the same symmetry (even modes), and consequently the Zak phase is 0. 
Therefore, the Zak phase of the second band is altered after the band crossing, which indicates the topological phase transition. 

The Zak phase of isolated passbands is marked in magenta in Fig.~\ref{Fig3}. According to Figs.~\ref{Fig3}(d) and (f), we can see that for Configurations S1 and S2, the Zak phase of the first band is 0; that S1 and S2 have an overlap part in the second BG frequency range; and that when the soft dielectric PC plate turns from  S1 to S2, the Zak phase of the second passband transitions from $\pi$ to 0. 

In passing, we recall that the Zak phase of the $j$th band can also be calculated by the following expression \citep{xiao2015geometric,yin2018band},
\begin{equation} \label{56}
\theta _{j}^{\textrm{Zak}}=\int_{-\pi /l}^{\pi /l}{\left[ \operatorname{i}\int\limits_{\text{unit cell}}{\dfrac{1}{2\rho c^2}\textrm{d}x_1\xi_{j,k}^{*}\left( x_1 \right){{\partial }_{k}}{{\xi}_{j,k}}\left(x_1 \right)} \right]}\textrm{d}k,
\end{equation}
where ${1}/{2\rho c^2}$ is the weight function of this system with $\rho$ and $c$ being mass density and bending wave velocity in the current deformed configuration, respectively; ${{\xi}_{j,k}}\left(x_1 \right) = {{{u_2}}_{j,k}}\left( x_1 \right){{{\text{e}}}^{-\text{i}kx_1}}$ is the normalized cell-periodic Bloch displacement eigenfunction for the $j$th isolated band and Bloch wave number $k$.
After obtaining the Zak phase, the topological property of any BG can be determined by the BG sign $\varsigma$, which is expressed as \citep{xiao2014surface}
\begin{equation} \label{57}
{\mathop{\rm sgn}} \left[ {{\varsigma ^{\left( j \right)}}} \right] = {\left( { - 1} \right)^j}{\left( { - 1} \right)^r}{\rm{exp}}\left( {{\rm{i}}\sum\limits_{\beta  = 1}^{j} {\theta _\beta ^{\text{Zak}}} } \right),
\end{equation}
where $r$ is an integer indicating the number of band crossing points beneath the $j$th BG. Based on Eq.~\eqref{57}, the second BG signs with $\varsigma < 0$ and $\varsigma >0$ are marked in Fig.~\ref{Fig3} by the green and yellow stripes, respectively, to demonstrate different BG topological properties.

Moreover, the transition of the BG topological property can also be verified by the eigenmode exchange of the BG edge states (see Figs.~\ref{Fig4}(b), (c), (e) and (f)). To be specific, edge states $B_1$ and $C_2$ in Figs.~\ref{Fig4}(b) and (f) indicate the same antisymmetric distribution of displacement fields with respect to the center of unit cell, while the same symmetric edge states $B_2$ and $C_1$ are shown in Figs.~\ref{Fig4}(c) and (e). 
Therefore, the second BG of Configuration S1 is topologically different from that of Configuration S2.
It is well known that a topological state exists at the interface of a mixed PC waveguide made of two PC elements possessing overlapped BGs with different topological properties. 
Hence, the topological interface state is feasible in a finite soft dielectric PC waveguide consisting of S1 and S2 unit cells, as we display next.

\begin{figure}[htbp]
	\centering
	\setlength{\abovecaptionskip}{5pt}	
	\includegraphics[width=0.7\textwidth]{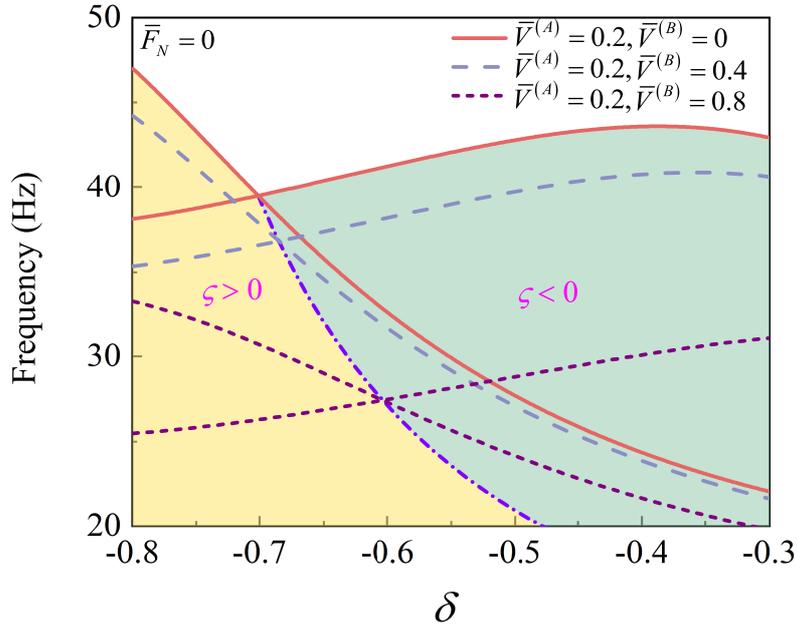}
	\caption{Topological phase diagram of an ideal dieletric Gent PC plate ($J_m=10$) in the absence of axial force. Three groups of solid, dashed and densely dashed curves stand for the frequency limits of the second BG as a function of the structural parameter $\delta$ under the electric voltages ${{\overline{V}}^{(A)}}=0.2$, ${\overline{V}}^{(B)}=0$; ${{\overline{V}}^{(A)}}=0.2$, ${\overline{V}}^{(B)}=0.4$; and ${{\overline{V}}^{(A)}}=0.2$, ${\overline{V}}^{(B)}=0.8$. The frequencies of topological transition points for these three groups of curves are 39.5, 36.8 and 27.4 Hz, respectively. The dash-dotted line is the topological phase curve when the electric voltage ${{\overline{V}}^{(A)}}$ is fixed and ${{\overline{V}}^{(B)}}$ varies from 0 to 1, where the topological transition points are located. The yellow and green filled regions are delimited by the topological phase line, and indicate the second BGs with $\varsigma > 0$ and $\varsigma <0$, respectively.}
	\label{Fig5}
\end{figure}

Now we discuss the effect of the electric voltage on the band structure, especially the frequency limits of the second BG. 
Fig.~\ref{Fig5} shows a topological phase diagram, which illustrates the topological transition process (i.e. the variation of frequency limits of the second BG as a function of $\delta$) for the ideal dielectric Gent PC plate ($J_m=10$) at different electric voltages with $\overline F_N=0$. 
We see from Fig.~\ref{Fig2} that the thinner PC sub-plate $A$ is more sensitive than the thicker sub-plate $B$ to the electric voltage. For a broad tunable range of electric voltage applied to sub-plate $B$, the electric voltage value applied to sub-plate $A$ is set to be a fixed value as ${{\overline{V}}^{(A)}}=0.2$ such that the overlap in the second BG for different configurations can be reserved. Meanwhile, with the increase of the electric voltage applied to sub-plate $B$ from 0 to 0.8, Fig.~\ref{Fig5} shows that the position of the topological transition point is shifted down, and that the value of $\delta$ where the BG closes becomes larger. 
For example, for ${\overline{V}}^{(B)}=0$, 0.4 and 0.8, the corresponding topological transition points are located at $\delta=-0.701$, $-0.685$, and $-0.604$, with their frequencies being 39.5, 36.8 and 27.4 Hz, respectively. The dash-dotted line in Fig.~\ref{Fig5} is the topological phase curve, which is connected by the topological transition points for different electric voltages ${\overline{V}}^{(B)}$ varying from 0 to 1 with a fixed ${\overline{V}}^{(A)}=0.2$. The topological phase diagram is delimited by this topological phase curve, and the divided two regions with different topological properties can be obtained. 
Moreover, referring to the topological phase diagram, we can design finite PC waveguide composed of elements with distinct toplogical properties, and acquire the topological interface states with tunable working frequency.
\begin{figure}[h!]
	\centering
	\setlength{\abovecaptionskip}{5pt}
	\setlength{\belowcaptionskip}{0pt}
	\includegraphics[width=0.85\textwidth]{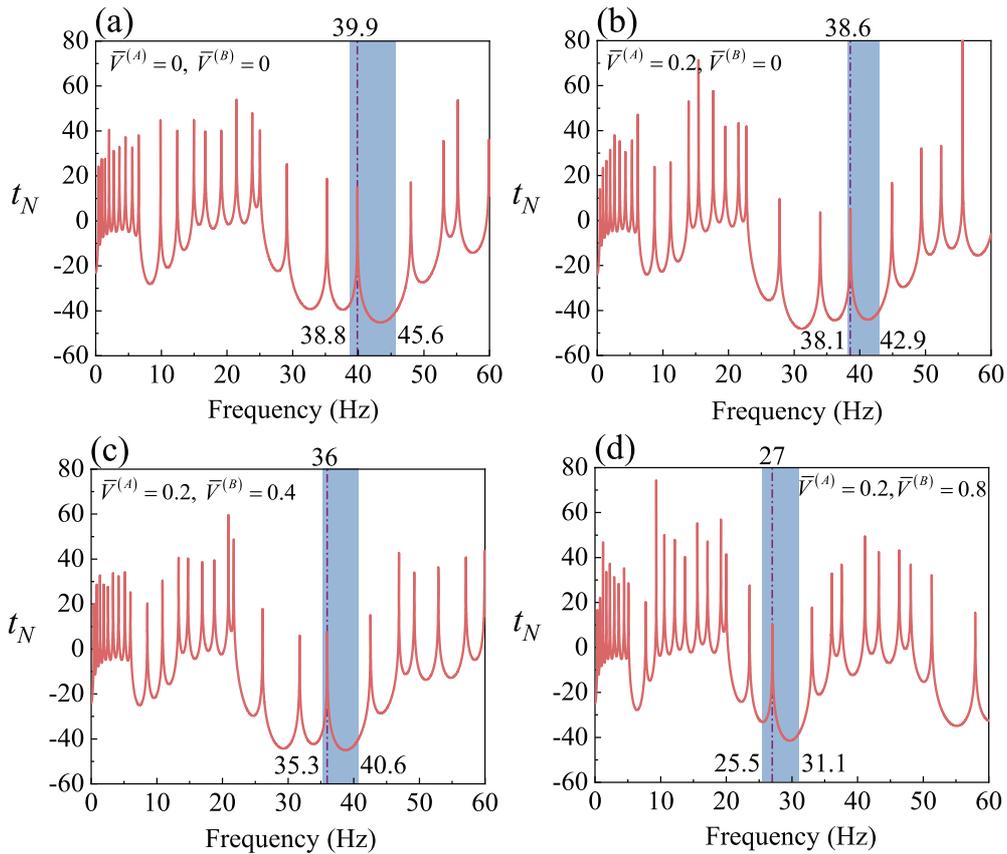}
	\caption{Transmission spectra of a mixed finite Gent ($J_m=10$) dielectric PC plate waveguide composed of five S1 unit cells ($\delta=-0.3$) and five S2 unit cells ($\delta=-0.8$) for different electric voltages: (a) $\overline V^{(A)}=0$, $\overline V^{(B)}=0$; (b) $\overline V^{(A)}=0.2$, $\overline V^{(B)}=0$; (c) $\overline V^{(A)}=0.2$, $\overline V^{(B)}=0.4$; (d) $\overline V^{(A)}=0.2$, $\overline V^{(B)}=0.8$. The electric voltages applied to S1 and S2 unit cells are the same. The frequency limits of the overlapped second BG and the frequencies of topological interface states are indicated in the figure.}
	\label{Fig6}
\end{figure}

Now, Eq.~\eqref{54} is employed to calculate the transmission behaviors (propagation from the left end to the right end) of a mixed finite PC plate waveguide in order to examine the existence and tunability of the topological interface state by the electric voltage. Accordingly, a mixed finite PC plate waveguide is designed here, which consists of five S1 unit cells ($\delta=-0.3$) connecting five S2 unit cells ($\delta=-0.8$). Fig.~\ref{Fig6} displays the theoretical results for the mixed PC plate waveguide subjected to different electric voltages. 
Note that although $\delta$ has different values, the electric voltages applied to S1 and S2 unit cells are the same. The frequency limits of the overlapped second BG for the two PC elements with different topological properties, and the frequencies of topological interface states (marked by the purple dash-dotted line) of the mixed finite PC plate are labelled at the bottom and top of the corresponding sub-figures, respectively. We can observe that the transmission peaks occur in the overlapped BG range (marked in blue) in Figs.~\ref{Fig6}(a)-(d). Compared with the result for the natural undeformed configuration (see Fig.~\ref{Fig6}(a)), the overlapped frequency range in the second BG is kept and the topological interface state can be tuned to a lower frequency with the application of electric voltage. For instance, the overlapped BG varies from (38.8 Hz, 45.6 Hz) for  $\overline V^{(A)}=0$ and $\overline V^{(B)}=0$ to (25.5 Hz, 31.1 Hz) for $\overline V^{(A)}=0.2$ and $\overline V^{(B)}=0.8$ with the frequency of topological interface state tuned from 39.9 Hz to 27 Hz.

\begin{figure}[t]
	\centering
	\setlength{\abovecaptionskip}{5pt}	
	\includegraphics[width=0.9\textwidth]{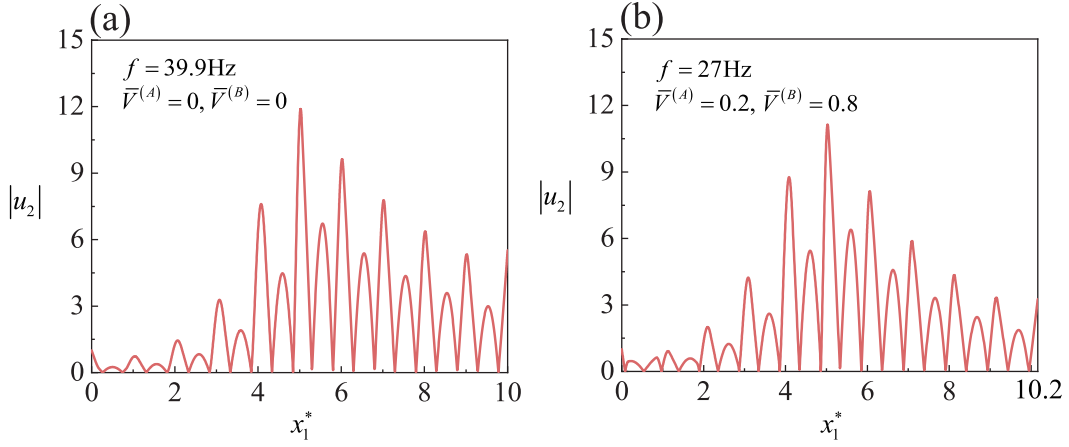}
	\caption{The amplitude of displacement $|u_2|$ of the mixed finite Gent PC plate waveguide as a function of the dimensionless axial coordinate ${x_1^{*}} = {x_1}/{l}$ ($l$ is the length of the deformed unit cell for S1 configuration) at peak frequencies 39.9 Hz ($\overline V^{(A)}=0$, $\overline V^{(B)}=0$) (a)  and 27 Hz  ($\overline V^{(A)}=0.2$, $\overline V^{(B)}=0.8$) (b) in absence of the axial force ($\overline F_N=0$).}
	\label{Figdisp}
\end{figure}

Furthermore, we plot in Fig.~\ref{Figdisp} the displacement distribution of the mixed finite Gent PC plate waveguide at the frequencies of transmission peaks for the unloaded and loaded cases ($\overline V^{(A)}=0.2$, $\overline V^{(B)}=0.8$). Figs.~\ref{Figdisp}(a) and (b) display the amplitude of displacement  $|u_2|$ at the transmission peak frequencies 39.9 Hz ($\overline V^{(A)}=0$, $\overline V^{(B)}=0$) and 27 Hz ($\overline V^{(A)}=0.2$, $\overline V^{(B)}=0.8$) as a function of the normalized axial coordinate ${x_1^{*}} = {x_1}/{l}$, where $l$ is the length of the deformed unit cell for S1 Configuration. For the peak frequencies of topological interface states, the displacement amplitudes at the interfaces delimiting different unit cells of the mixed PC plate waveguide are illustrated in Fig.~\ref{Figdisp}. We observe that the displacement mode concentrates at the interface between two PC plate elements and decays rapidly to the two ends, which is an obvious sign of topological interface state. Specifically, the amplification of the displacement amplitude at the interface is over 10 times larger than the input signal. Different from other transmission peaks ascribed to the resonance of finite structure \citep{chen2019tunable}, the peaks corresponding to the topological interface state are not affected by the excitation location and boundary conditions \citep{yin2018band,chen2021low}. This is a result of the robust mechanism based on the conflict of topological property.

\begin{figure}[htbp]
	\centering
	\setlength{\abovecaptionskip}{5pt}	
	\includegraphics[width=0.9\textwidth]{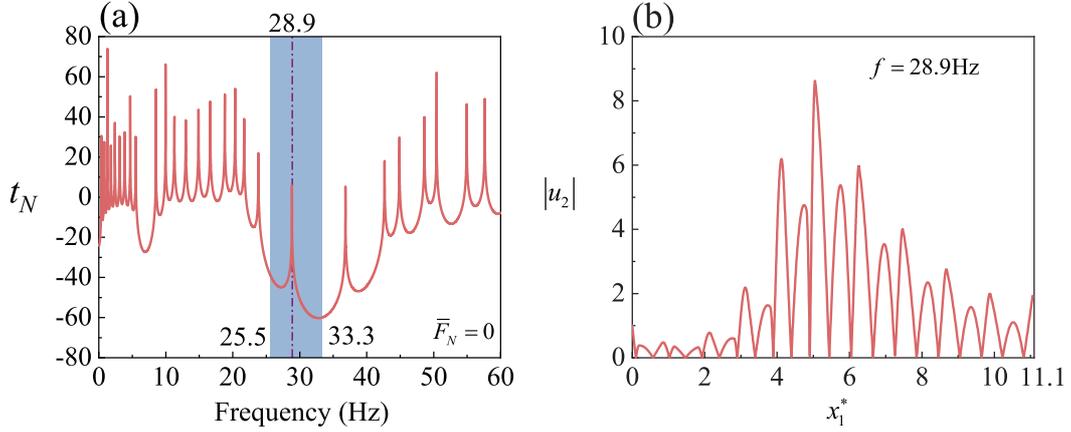}
	\caption{(a) Transmission spectrum of a mixed finite Gent ($J_m=10$) dielectric PC plate waveguide composed of five  S1 unit cells and five S2 unit cells when $\overline F_N=0$. The electric voltages applied to the two PC elements are different:  $\overline V^{(A)}=0.2$, $\overline V^{(B)}=0$ for Configuration S1; $\overline V^{(A)}=0.2$, $\overline V^{(B)}=0.8$ for Configuration S2. (b) The amplitude of displacement $|u_2|$ of the mixed finite Gent PC plate waveguide as a function of the dimensionless axial coordinate ${x_1^{*}} = {x_1}/{l}$ ($l$ is the length of the deformed unit cell for S1) at the peak frequency 28.9 Hz.}
	\label{Fig11}
\end{figure}

The results shown above are limited to applying the same electric loads on two different PC elements of the mixed finite PC waveguide. As we can see from Fig.~\ref{Fig6}, the overlapped BG may become narrower when the electric voltage varies, which is not beneficial for the existence and tunability of the topological interface state. We can see from Fig.~\ref{Fig5} that the second BG for different PC elements (at two sides of the topological phase curve) may have a broader overlapped part with opposite topological properties when applying different electric loads on the two PC elements of the mixed finite PC waveguide. Fig.~\ref{Fig11}(a) shows the transmission spectrum of the mixed finite PC plate waveguide with $\overline V^{(A)}=0.2$, $\overline V^{(B)}=0$ applied to S1 and $\overline V^{(A)}=0.2$, $\overline V^{(B)}=0.8$ applied to S2, respectively. Compared with the case of applying the same electric voltages on two different PC elements, the overlapped BG obtained in Fig.~\ref{Fig11} is much wider. The topological interface state denoted by the  transmission peak in the overlapped BG is still observed and its displacement distribution at the peak frequency 28.9 Hz is plotted in Fig.~\ref{Fig11}(b). We can see that the displacement at the interface is over 8 times larger than the input signal and attenuates rapidly to the ends of the mixed PC plate waveguide.

Hence, with  flexible application of voltage and proper geometrical design of mixed PC waveguides, we can tune the frequency of topological interface state in a wide range.


\subsection{Tunable effect of the axial force on bending waves} \label{Sec5-3}

\begin{figure}[t]
	\centering
	\setlength{\abovecaptionskip}{5pt}	
	\includegraphics[width=0.9\textwidth]{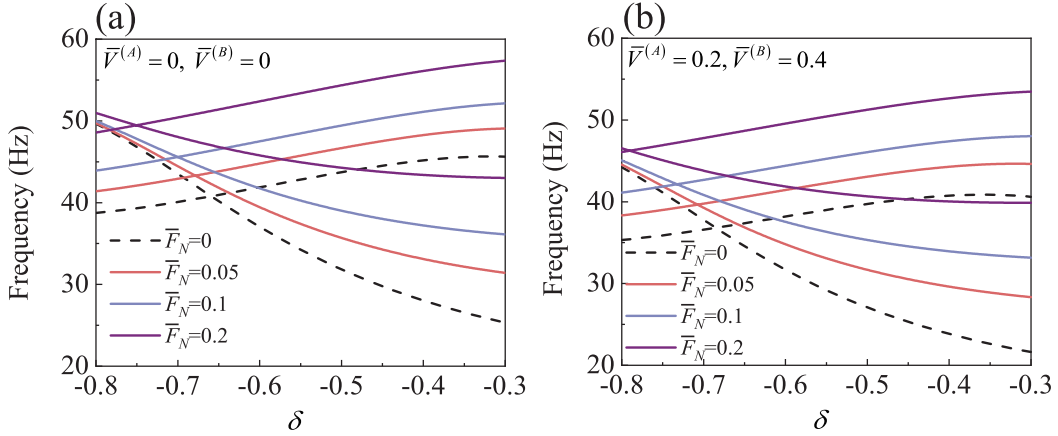}
	\caption{Frequency limits of the second BG as a  function of the structural parameter $\delta$ in the soft dielectric Gent PC plate ($J_m=10$) under the action of different axial forces $\overline F_N=0$, 0.05, 0.1 and 0.2, with (a) $\overline V^{(A)}=0$, $\overline V^{(B)}=0$ and (b) $\overline V^{(A)}=0.2$, $\overline V^{(B)}=0.4$.}
	\label{Fig7}
\end{figure}

In addition to changing the electric voltage, the axial force can also be applied to steer the topological interface state in the soft dielectric PC plate waveguide. 
For different axial forces $\overline F_N=0$, 0.05, 0.1 and 0.2, we highlight in Figs.~\ref{Fig7}(a) and (b) the variation of the frequency limits of the second BG with the structural parameter $\delta$ for the dielectric Gent PC plate ($J_m=10$) with $\overline V^{(A)}=0$, $\overline V^{(B)}=0$ and  $\overline V^{(A)}=0.2$, $\overline V^{(B)}=0.4$, respectively. 

For $\overline V^{(A)}=0$ and $\overline V^{(B)}=0$, four groups of band inversion curves in Fig.~\ref{Fig7}(a) reveal that the axial force affects the topological transition in a different way from the electric voltage. 
Hence, an increase in axial force may result in an increase in frequency and a decrease of structural parameter $\delta$ corresponding to the topological transition point, which is in contrast to the case by increasing the voltage (see Fig.~\ref{Fig5}). 
The underlying mechanism can be explained as follows. 
For the dielectric PC plate subjected to the electric load only, the resultant deformation in Fig.~\ref{Fig5} has not reached the strain-stiffening stage. 
In this case, the change in effective modulus ${\cal A}_0^{\rm{e}} $ cannot cancel the influence of the increase of geometric size on the frequency. 
Consequently, the increase of electric voltage yields the decrease in the frequency of topological interface state. 

For the PC plate subjected to a varying axial force with fixed electric voltage, the axial force not only increases the effective modulus ${\cal A}_0^{\rm{e}} $, but also contributes to the bending moment of the plate (see Eq.~\eqref{32}), and their influence on the BG of PC plate is a more dominant factor than the change in geometry. 
Hence, with the increase of axial force, the working frequency of bending waves is moved upwards.
For example, for the undeformed PC plate ($\overline F_N=0$), the frequency of topological transition point is $f=40.7 Hz$, with $\delta = -0.66$, wheras for $\overline F_N=0.1$, the values are $f= 45.6$ Hz and $\delta = -0.7$. 
When the PC plate is subjected to the voltages $\overline V^{(A)}=0.2$ and $\overline V^{(B)}=0.4$, similar observations can be made from Fig.~\ref{Fig7}(b) through adjusting the axial force. 
\begin{figure}[h!]
	\centering
	\setlength{\abovecaptionskip}{5pt}	
	\includegraphics[width=1.0\textwidth]{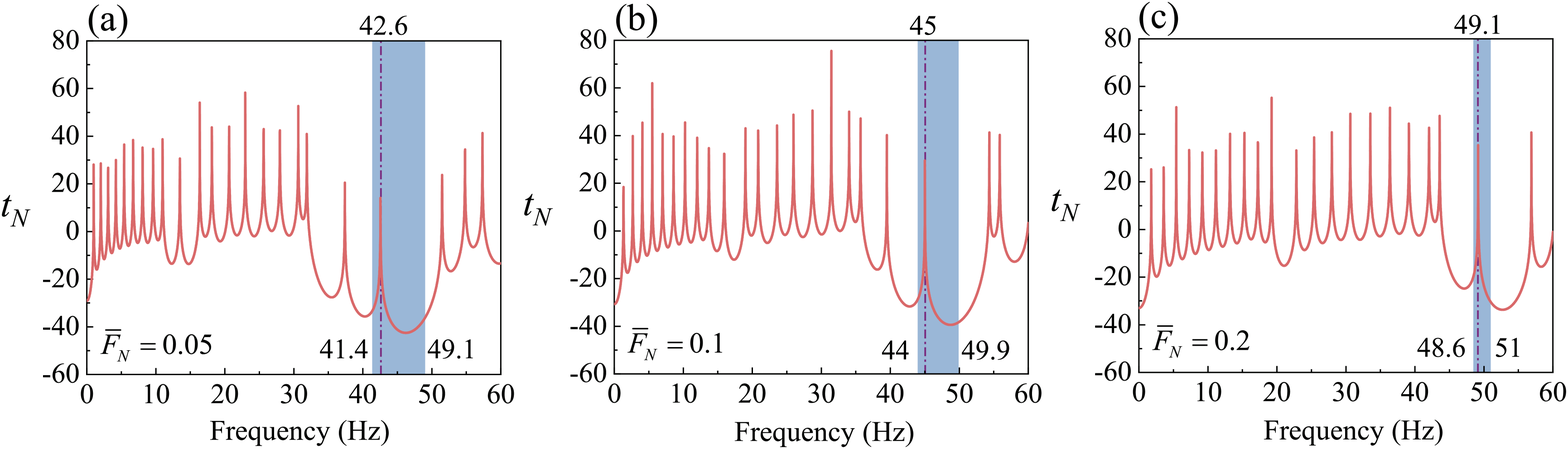}
	\caption{Transmission spectra of a mixed finite Gent ($J_m=10$) PC plate waveguide composed of five S1 unit cells ($\delta=-0.3$) and five S2 unit cells ($\delta=-0.8$) for $\overline V^{(A)}=\overline V^{(B)}=0$ and different axial forces: (a) $\overline F_N=0.05$; (b) $\overline F_N=0.1$; (c) $\overline F_N=0.2$.}
	\label{Fig8}
\end{figure}
\begin{figure}[htbp]
	\centering
	\setlength{\abovecaptionskip}{5pt}	
	\includegraphics[width=1.0\textwidth]{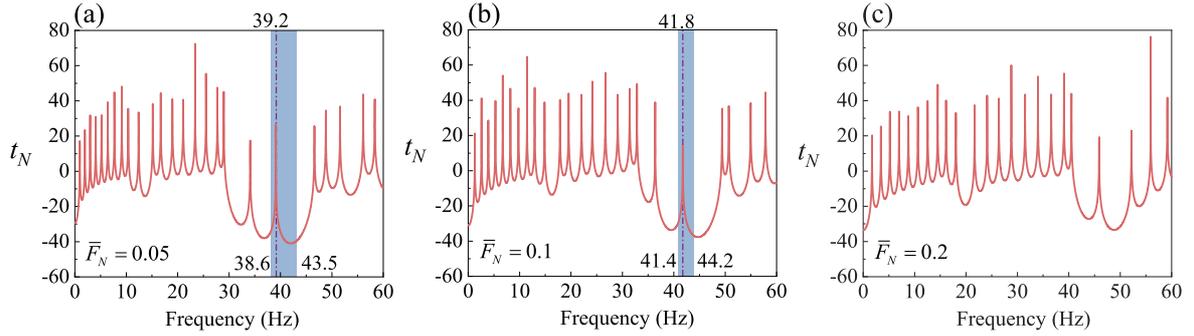}
	\caption{Transmission spectra of a mixed finite Gent ($J_m=10$) PC plate waveguide composed of five S1 unit cells ($\delta=-0.3$) and five S3 unit cells ($\delta=-0.78$) for $\overline V^{(A)}=0.2$, $\overline V^{(B)}=0.4$ and different axial forces: (a) $\overline F_N=0.05$; (b) $\overline F_N=0.1$; (c) $\overline F_N=0.2$.}
	\label{Fig9}
\end{figure}

Next, we present the transmission behaviors of the mixed finite PC plate waveguide for a prescribed electric load and different values of the axial force. 
In the case  $\overline V^{(A)}=\overline V^{(B)}=0$, Fig.~\ref{Fig8} displays the transmission spectra of the mixed finite waveguide (composed of five S1-type unit cells and five S2-type unit cells) subjected to different axial forces $\overline F_N=0.05$, 0.1 and 0.2 (see Fig.~\ref{Fig6}(a) for $\overline F_N=0$). 
We see that the transmission peak standing for the topological interface state exists in the second overlapped BG for different values of axial force. 
Besides, with the increase in the axial force, the position of the peak frequency is moved upwards, although the width of the overlapped BG narrows. For example, for $\overline F_N=0.05$, the common frequency range of the second BG is (41.4 Hz, 49.1 Hz) with the peak frequency being 42.6 Hz, while for $\overline F_N=0.2$, the corresponding overlapped BG and peak frequencies become (48.6 Hz, 51 Hz) and 49.1 Hz, respectively.

It is also worth noting from Fig.~\ref{Fig7}(b) that for $\overline F_N=0.2$,
the topological transition point is located at  $\delta=-0.79$ and $f=46.3$ Hz, which is in the vicinity of $\delta=-0.8$. 
With an appropriate structural design, the existence of a topological interface state can be controlled by adjusting the axial force. 
Here, we name the configuration with $\delta=-0.78$ as Configuration S3.
Similar to Fig.~\ref{Fig8}, Fig.~\ref{Fig9} illustrates the corresponding transmission curves for the mixed finite PC plate composed of five S1 unit cells and five S3 unit cells for the fixed voltage $\overline V^{(A)}=0.2$, $\overline V^{(B)}=0.4$ and different axial forces 0.05, 0.1 and 0.2. 
Analogous observations for the tunability of the overlapped BG frequency and peak frequency under the axial force to those in Fig.~\ref{Fig8} can be made. 

Recall that the existence of topological interface state requires an overlapped BG with different topological properties. 
When the axial force increases from 0 to 0.2, the topological properties of Configurations S1 and S3 change from being different to being the same (see Fig.~\ref{Fig7}(b)). 
Under this circumstance, there is no topological interface state in the second BG in Fig.~\ref{Fig9}(c) for $\overline F_N=0.2$. Note that other transmission peaks can be observed in Fig.~\ref{Fig9}(c), but they are resonance peaks relying on excitation and boundary conditions.
  
  
\section{Conclusions}\label{section6}
 
 
In this work, we investigated incremental bending wave motions superimposed on finite static deformations in 1D incompressible dielectric PC plates under the combined action of axial force and electric voltage. Specifically, we discussed the topological interface state and its tunability via the mechanical and electric loads in the finite PC plate waveguide consisting of two types of PC elements with overlapped BG and different topological properties. 

First, we employed the theoretical framework of nonlinear electroelasticity to determine the nonlinear deformation of the dielectric PC plate with step-wise cross-sections. Furthermore, we employed a counterpart of the pre-stressed Euler-Bernoulli beam model to describe the incremental bending wave motions in the soft dielectric PC plate. 
We used the Spectral Element Method to derive the dispersion relation of the infinite PC plate and transmission coefficient of the finite PC plate waveguide. 
Finally, we presented numerical results in detail to demonstrate the effects of axial force and electric voltage on the position (i.e. structural parameter) of the topological transition point and on the frequency of the topological interface state. Several useful conclusions based on the numerical results are listed as follows:  
 	
 \begin{enumerate}[(1)]
    \item The topological transition process can be observed when the initial structural parameter varies, and we can apply external mechanical and electric loads to tune the position of the transition point. 
    
 	\item For the finite PC plate waveguide under a prescribed axial force, the frequency of the topological interface state decreases with an increase in electric voltage. 
 	
 	\item For the finite PC plate waveguide under a fixed electric voltage, increasing the axial force raises the frequency of the topological interface state. When the axial force reaches a large value, the topological interface state may disappear because the requirement of two PC elements with overlapped and topologically different BGs is not satisfied any more.
 	
 	\item Based on the topological phase diagram, the topological interface state can be adjusted in a wide range by applying different electric voltages separately on two PC elements in a well-designed PC waveguide system.
 	
 \end{enumerate}	

The numerical results indicate that the axial force and electric voltage are effective methods to actively control the topological interface state in the 1D soft dielectric PC plate waveguide operating at a low working frequency. Our study lays down a solid theoretical foundation for the design of soft dielectric topological PC devices in order to cater to  real-world  applications such as low-pass filters, high-sensitivity biomedical detectors, and tunable energy harvesters.



\section*{Acknowledgements}

This work was supported by the National Natural Science Foundation of China (Nos. 11872329, 12192210, and 12192211),  the 111 Project (No. B21034), the Natural Science Foundation of Zhejiang Province (No. LD21A020001), and the China Scholarship Council (CSC). WB gratefully acknowledges the support of the European Union Horizon 2020 Research and Innovation Programme under the Marie Sk{\l}odowska-Curie Actions, Grant No.~896229.
  
\appendix



\section*{References}

\bibliographystyle{elsarticle-harv.bst}
\nocite{*}
\bibliography{topo1D.bib}







\end{document}